\documentclass[useAMS,onecolumn]{mn2e}
\usepackage{graphicx}
\usepackage{times}
\usepackage{amsmath}
\title {Early Afterglows in Wind Environments Revisited}

\author[Y. C. Zou, X. F. Wu, and Z. G. Dai]{Y. C. Zou, X. F. Wu, and Z. G. Dai \thanks{E-mail: zouyc@nju.edu.cn(YCZ); xfwu@nju.edu.cn(XFW); dzg@nju.edu.cn(ZGD) }
\\
Department of Astronomy, Nanjing University, Nanjing 210093, China\\
 }
\begin{document}

\date{\today}

\pagerange{\pageref{firstpage}--\pageref{lastpage}} \pubyear{2005}
\maketitle
\label{firstpage}

\begin{abstract}
When a cold shell sweeps up the ambient medium, a forward shock and a reverse shock will form. We analyze the reverse-forward shocks in a wind environment, including their dynamics and emission. An early afterglow is emitted from the shocked shell, e.g., an optical flash may emerge. The reverse shock behaves differently in two approximations: relativistic and Newtonian cases, which depend on the parameters, e.g., the initial Lorentz factor of the ejecta. If the initial Lorentz factor is much less than $114 E_{53}^{1/4} \Delta_{0,12}^{-1/4} A_{*,-1}^{-1/4}$, the early reverse shock is Newtonian. This may take place for the wider of a two-component jet, an orphan afterglow caused by a low initial Lorentz factor, and so on. The synchrotron self absorption effect is significant especially for the Newtonian reverse shock case, since the absorption frequency $\nu_a$ is larger than the cooling frequency $\nu_c$ and the minimum synchrotron frequency $\nu_m$ for typical parameters. For the optical to X-ray band, the flux is nearly unchanged with time during the early period, which may be a diagnostic for the low initial Lorentz factor of the ejecta in a wind environment. We also investigate the early light curves with different wind densities, and compare them with these in the ISM model.
\end{abstract}


\section{\label{sec1:level1}Introduction}

Long-duration gamma-ray bursts (GRBs) may originate from the collapse of massive stars (Woosley 1993; Paczy\'nski 1998). The probable associations between GRBs and supernovae have been detected in several cases, e.g., the most confirmed GRB 980425 / SN 1998bw (Galama et~al. 1998, Kulkarni et al. 1998) and GRB 030329 / SN 2003dh (Hjorth et~al. 2003), which give a firm link to the collapsar model. The light curves of several afterglows also show the supernova component, such as GRB 970228 (Reichart 1999), 980326 (Bloom et~al. 1999), 011121 (Bloom et~al. 2002, Greiner et~al. 2003), 021004 (Schaefer et~al. 2003), and so on. Thus, the surrounding environment is wind-type. Much work about wind-type environment analyses has been done (Dai \& Lu 1998; M\'esz\'aros, Rees \& Wijers 1998; Panaitescu \& Kumar 2000; Chevalier \& Li 2000), including the features of the afterglow light curves and the comparison with the interstellar medium (ISM) model.

At the beginning of the interaction between the ejected shell and environment, an early afterglow will emerge from the reverse shock, as predicted by M\'esz\'aros \& Rees (1997) and Sari \& Piran (1999b). The prompt optical emission from GRB 990123 (Akerlof et~al. 1999) and GRB 021211 (Fox et~al. 2003; Li et~al. 2003) were observed, though observations of a very early afterglow were difficult before the {\it{Swift}}'s launch (Gehrels et~al. 2004). The optical flash of GRB 990123 was immediately analyzed (M\'esz\'aros \& Rees 1999; Sari \& Piran 1999a). They pointed out that the optical flashes mainly come from the contribution of the reverse shock. Then, plenty of theoretical analyses were advanced. The dynamics, numerical results of optical and radio emission, and analytical light curves, from the reverse and forward shocks in uniform environments were discussed (Sari \& Piran 1995; Kobayashi \& Sari 2000; Kobayashi 2000). Early afterglows in wind environments were also considered by several groups (Chevalier \& Li 2000; Wu et~al. 2003; Kobayashi \& Zhang 2003). Panaitescu \& Kumar (2004) considered the reverse-forward shock scenario and the wind bubble scenario for the two observed optical flashes from GRB 990123 and GRB 021211. However, little discussion comes to a Newtonian reverse shock in wind environment. We consider this case in the following sections.

Kobayashi, M\'esz\'aros \& Zhang (2004) noticed the synchrotron self-absorption (SSA) effect on the early afterglow. We find that other parameters like the thickness of the shell and the initial isotropic kinetic energy can also influence the self absorption, even up to the optical wavelength. In this paper, we derive the complete scaling-laws of the SSA frequency for all cases. 

In the simulations of Zhang, Woosley \& MacFadyen (2003), the initial Lorentz factor can be as low as about tens. For the structured jet model (Kumar \& Granot 2003), it is likely that the jet has low Lorentz factors at the wings of the jet. Huang et~al. (2002) considered that a jet with an initial Lorentz factor less than 50 may cause an orphan afterglow. Rhoads (2003) also pointed out that the fireball with a low initial Lorentz factor will produce a detectable afterglow, though no gamma-ray emission is detectable. There are indications that some GRBs ejecta have two components: a narrow ultra-relativistic inner core, and a wide mildly relativistic outer wing (Berger et~al. 2003; Huang et~al. 2004; Wu et~al. 2005; Peng et~al. 2005). When the mildly relativistic shell collides with the wind environment, the reverse shock is Newtonian. Kobayashi (2000) considered the Newtonian reverse shock in a uniform environment. However, no systematic analysis has come into the Newtonian reverse shock in a wind environment. In this work, we discuss the Newtonian reverse shock, which is mainly caused by a low initial Lorentz factor. In this case, the optical emission flux from the Newtonian shocked region exceeds that from the relativistic forward shocked region.

Some authors have used the early afterglow as a diagnostic tool of gamma ray bursts' parameters for the ISM case (Zhang, Kobayashi \& M\'esz\'aros 2003; Nakar et~al. 2004) and for the wind case (Fan et~al. 2004; Fan et~al. 2005). Accordingly, the behavior of early reverse-forward shocks should be completely described for the wind case. We derive the analytical scaling-laws of dynamics and radiations for both relativistic and Newtonian reverse shock cases in the wind environment in \S\S \ref{sec:hydro} and \ref{sec:emission}, and give the numerical results of radio to X-ray light curves in \S \ref{sec:numerical-results}. We present some discussions in \S \ref{sec:conclusions}.

\section{Hydrodynamics}\label{sec:hydro}

Let's consider a uniform and cold relativistic coasting shell with isotropic kinetic energy $E_0$, Lorentz factor $\gamma_4=\eta+1 \gg 1$, and width in observer's frame $\Delta_0$, ejected from the progenitor of the GRB. This shell sweeps up a free wind environment with number density $n_1=Ar^{-2}$, where $\eta$ is the initial ratio of $E_0$ to the rest mass of the ejecta (Piran, Shemi \& Narayan 1993). The interaction between the shell and the wind develops a forward shock propagating into the wind and a reverse shock propagating into the shell. The two shocks separate the system into four regions: (1) the unshocked approximately stationary wind (called region 1 hereafter), (2) the shocked wind (region 2), (3) the shocked shell material (region 3), and (4) the unshocked shell material (region 4). By using the shock jump conditions (Blandford \& McKee 1976, BM hereafter) and assuming the equality of pressures and velocities beside the surface of the contact discontinuity, the values of the Lorentz factor $\gamma$, the pressure $p$, and the number density $n$ in the shocked regions can be estimated as functions of $n_1$, $n_4$, and $\eta$, where $n_4=E_0/(\eta 4\pi r^2 \gamma_4 \Delta_0 m_p c^2)$ is the comoving number density of region 4.

Analytical results can be obtained in both relativistic and Newtonian reverse shock limit. These two cases are divided by comparison between $f$ and $\gamma_4^2$, where $f\equiv n_4/n_1$ is the ratio of the number densities between the unshocked shell and the unshocked wind (Sari \& Piran 1995). As shown by Wu et~al. (2003) for the wind environment case, $f=l/(\eta^2\Delta_0)$, where $l = E_0/(4\pi A m_p c^2)$ is the Sedov length. If $f\gg \gamma_4^2$, the reverse shock is Newtonian (NRS), and if $f \ll \gamma_4^2$, the reverse shock is relativistic (RRS).

As discussed by Kobayashi \& Sari (2000), even for NRS, the adiabatic index of the post-shocked fluid can be taken as a constant $\hat{\gamma}=4/3$, because the electrons are still relativistic. Then the shock jump conditions can read (BM; Sari \& Piran 1995)
{\setlength\arraycolsep{2 pt}
\begin{eqnarray}
  e_2 / {n_2 m_p c^2} &=& \gamma_2 - 1,  n_2 / n_1 = 4\gamma_2+3,  \label{eq:jump1} \\
  e_3 / {n_3 m_p c^2} &=& \bar {\gamma_3} - 1,  n_3/n_4 = 4\bar{\gamma_3} +3,  \label{eq:jump2}
\end{eqnarray}}
where $m_p$ is the proton mass, $e_2$ and $e_3$ are the comoving energy densities of region 2 and region 3 respectively, and $n_2$ and $n_3$ are the corresponding comoving number densities of particles, which are assumed to consist of protons and electrons. The relative Lorentz factor between region 3 and region 4 is
\begin{equation}
  \bar{\gamma_3}=\gamma_3\gamma_4(1-\sqrt{1-1/\gamma_3^2}\sqrt{1-1/\gamma_4^2}).
  \label{eq:gamma_34}
\end{equation}
Assuming $\gamma_2 = \gamma_3$, and $\gamma_2, \gamma_4 \gg 1$, $\bar{\gamma_3}$ can be expressed as $\bar{\gamma_3}\simeq (\gamma_4/\gamma_2+\gamma_2/\gamma_4)/2$. The asymptotic solution is $\gamma_3 \simeq \frac{1}{\sqrt{2}} \gamma_4^{1/2}f^{1/4}$, $\bar{\gamma_3} \simeq \frac{1}{\sqrt{2}} \gamma_4^{1/2} f^{-1/4}$ for RRS, while $\gamma_3 \simeq \gamma_4$ and $\bar{\gamma_3}-1 \simeq \frac{4}{7}\gamma_4^2 f^{-1}$ for NRS.

The time it takes the reverse shock to cross the shell in the burster's frame is given by (Sari \& Piran 1995)
\begin{equation}
  t_\Delta=\frac{\Delta_0}{c(\beta_4-\beta_3)}\left(1-\frac{\gamma_4 n_4}{\gamma_3 n_3}\right).
\label{eq:t_delta}
\end{equation}
There are two simple limits involved in the problem: NRS and RRS, in which we can get analytical results. The relative Lorentz factor $\bar{\gamma_3}$ is constant in the whole reverse-shock period for RRS. $t_\Delta$ can be derived as $t_\Delta=\alpha \Delta_0 \gamma_4 f^{1/2}/c$, and the corresponding radius of the shell at time $t_\Delta$ is $r_\Delta \simeq c t_\Delta =\alpha \Delta_0 \gamma_4 f^{1/2} \simeq \alpha \sqrt{l \Delta_0} $, where the coefficient $\alpha=1/2$ for RRS and $\alpha=3/\sqrt{14}$ for NRS. We will discuss both cases separately in the following.

\subsection{Relativistic Reverse Shock Case}\label{sec:relat-reverse-shock}

In the RRS case, $f \ll \gamma_4^2$ (i.e., $\eta \gg 114 E_{53}^{1/4} \Delta_{0,12}^{-1/4} A_{*,-1}^{-1/4}$), using the relation between the observer's time and the radius $t_{\oplus}\simeq (1+z)r/2\gamma_3^2c$, where $z$ is the redshift of the GRB, we obtain $T\simeq (1+z) \Delta_0/2c \simeq 16.7 (1+z)\Delta_{0,12}$s as the RRS-crossing time in the observer's frame. We adopt the conventional denotation $Q = Q_{k}\times 10^{k}$ in this paper except for some special explanations. Using $e_2=e_3$, $\gamma_2=\gamma_3$, together with the above equations, we get the scaling-laws of the hydrodynamic variables for time $t_{\oplus}< T $,
{\setlength\arraycolsep{2 pt}
  \begin{eqnarray}
    \bar{\gamma_3} &\simeq& 1.9\eta_{2.5}E_{53}^{-1/4}\Delta_{0,12}^{1/4}A_{*,-1}^{1/4} \label{eq:RRS_g34}\\
    \gamma_3 &=& \gamma_2 \simeq 81.5 E_{53}^{1/4}\Delta_{0,12}^{-1/4}A_{*,-1}^{-1/4} \label{eq:RRS_g3}\\
    e_3 &=& e_2\simeq 1.4\times 10^{4} E_{53}^{-1/2}\Delta_{0,12}^{-3/2}A_{*,-1}^{3/2}\left(\frac{t_{\oplus}}{T}\right)^{-2} ~\mbox{erg cm}^{-3}\\
    N_{e,3} &\simeq& 2.1\times 10^{53} E_{53}\eta_{2.5}^{-1}\frac{t_{\oplus}}{T} \label{eq:RRS_e3}\\
    N_{e,2} &\simeq& 3.5\times 10^{51} E_{53}^{1/2}\Delta_{0,12}^{1/2}A_{*,-1}^{1/2}\frac{t_{\oplus}}{T}\label{eq:RRS_Ne2}
\end{eqnarray}}
where $A_{*}=3\times 10^{35} ~\mbox{cm}^{-1}$,  and $N_{e,i}$ is the number of electrons in the shocked region {\it{i}}. We note that $\gamma_3$ and $\bar{\gamma_3}$ do not depend on time. This is the property of wind environments, since the densities of the shell and the ambient environment have the same power-law relation with radius $r$ ($n \propto r^{-2}$).

After the reverse shock crosses the shell ($t_{\oplus}>T$), the shocked shell can be roughly described by the BM solution (Wu et~al. 2003; Kobayashi \& Zhang 2003; Kobayashi et~al. 2004),
{\setlength\arraycolsep{2 pt}
\begin{eqnarray}
  \gamma_3 &\propto& t_{\oplus}^{-3/8}, n_3 \propto t_{\oplus}^{-9/8}, e_3 \propto t_{\oplus}^{-3/2},
  r \propto t_{\oplus}^{1/4}, N_{e,3} \propto t_{\oplus}^0,\\
  \gamma_2 &\propto& t_{\oplus}^{-1/4}, n_2 \propto t_{\oplus}^{-5/4}, e_2 \propto t_{\oplus}^{-3/2},
  r \propto t_{\oplus}^{1/2}, N_{e,2} \propto t_{\oplus}^{1/2}.
\end{eqnarray}}
These variables can be scaled to the initial values ($t_{\oplus} = T$), which are given by the expressions for the time $t_{\oplus} < T$.

\subsection{Newtonian Reverse Shock Case}\label{sec:Newton-reverse-shock}

In the NRS case, $f\gg \gamma_4^2$, the time for the reverse shock crossing the shell is $T'\simeq t_{\eta} \simeq 2.9\times 10^3 (1+z) E_{53}\eta_{1.5}^{-4}A_{*,-1}^{-1}$s in the observer's frame, if we consider the spreading of the cold shell (Piran et~al. 1993). The evolution of the hydrodynamic variables before the time $T'$ are
{\setlength\arraycolsep{2 pt}
\begin{eqnarray}
  \bar{\gamma_3}-1 &\simeq& 0.57 \frac{t_{\oplus}}{T'} \label{eq:NRS_g34}\\
  \gamma_3 &=& \gamma_2 \simeq \gamma_4 \label{eq:NRS_g3}\\
  e_3 &=& e_2 \simeq 5.8 E_{53}^{-2} \eta_{1.5}^{6} A_{*,-1}^3 \left(\frac{t_{\oplus}}{T'}\right)^{-2} ~\mbox{erg cm}^{-3} \label{eq:NRS_e3}\\
  N_{e,3} &=& 2.1 \times 10^{54} E_{53} \eta_{1.5}^{-1} \left(\frac{t_{\oplus}}{T'}\right)^{1/2} \label{eq:NRS_Ne3}\\
  N_{e,2} &\simeq& 6.6\times 10^{52} E_{53} \eta_{1.5}^{-2} \frac{t_{\oplus}}{T'}\label{eq:NRS_Ne2}
\end{eqnarray}}
What should be noted is that the values for NRS are not suitable for mildly relativistic reverse shock case. Nakar \& Piran (2004) showed the difference between the approximated analytical solution and the numerical results in the case of uniform environments. And for the spreading of the shell, $f$ decreases with radius. At the crossing time, $\bar{\gamma_3} \simeq 1.57$ (see equation~(\ref{eq:NRS_g34})), which deviates from the Newtonian reverse shock approximation. More accurate values should be calculated numerically.

After the NRS crosses the shell, the Lorentz factor of the shocked shell can be assumed to be a general power-law relation $\gamma_3 \propto r^{-g}$ (M\'eszar\'os \& Rees 1999; Kabayashi \& Sari 2000). However, the forward shock is still relativistic, and can be described by the BM solution. The dynamic behavior is the same as the one in the RRS case. The scaling-law of the two regions are 
{\setlength\arraycolsep{2 pt}
\begin{eqnarray}
  \gamma_3 &\propto& t_{\oplus}^{-g/(1+2g)}, n_3 \propto t_{\oplus}^{-6(3+g)/7(1+2g)}, \nonumber \\ 
  e_3 &\propto& t_{\oplus}^{-8(3+g)/7(1+2g)},
  r \propto t_{\oplus}^{1/(1+2g)}, N_{e,3} \propto t_{\oplus}^0,\\
  \gamma_2 &\propto& t_{\oplus}^{-1/4}, n_2 \propto t_{\oplus}^{-5/4}, \nonumber \\
  e_2 &\propto& t_{\oplus}^{-3/2},
  r \propto t_{\oplus}^{1/2}, N_{e,2} \propto t_{\oplus}^{1/2}.
\end{eqnarray}}

\section{Emission}\label{sec:emission}

We now consider the synchrotron emission from the shocked material of region 2 and region 3. The shocks accelerate the electrons into a power-law distribution: $N(\gamma_e)d\gamma_e=N_\gamma \gamma_e^{-p}d\gamma_e (\gamma_e>\gamma_m)$, where $\gamma_m$ is the minimum Lorentz factor of the accelerated electrons. Assuming that constant fractions $\epsilon_e$ and $\epsilon_B$ of the internal energy go into the electrons and the magnetic field, we have $B=\sqrt{8\pi\epsilon_B e_i}$, where $e_i$ is the internal energy density of the shocked material. Regarding that the comoving internal energy of the electrons can also be written as $\epsilon_e e=\int_{\gamma_m}^\infty N_\gamma \gamma_e^{-p} \gamma_e m_e c^2d\gamma_e$, and the comoving number density $n=\int_{\gamma_m}^{\infty} N_\gamma \gamma_e^{-p}d\gamma_e$, one can get $\gamma_m = \epsilon_e (\bar{\gamma}-1)(m_p/m_e)(p-2)/(p-1)$, where $\bar{\gamma} = \bar{\gamma_3}$ or $\gamma_2$ corresponds to the reverse or forward shock, and $N_\gamma=n (p-1) \gamma_m^{p-1}$.

The cooling Lorentz factor $\gamma_c$ is defined when the electrons with $\gamma_c$ approximately radiate all their kinetic energy in the dynamical time, i.e., $(\gamma_c-1)m_ec^2=P(\gamma_c)t_{co}$, where $P(\gamma_e)=(4/3)\sigma_Tc(\gamma_e^2-1)(B^2/8\pi)$ (Rybicki \& Lightman 1979) is the synchrotron radiation power of an electron with Lorentz factor $\gamma_e$ in the magnetic field $B$, $t_{co}$ is the dynamical time in comoving frame (Sari et~al. 1998; Panaitescu \& Kumar 2000). Then the cooling Lorentz factor $\gamma_c=6\pi m_e c/(\sigma_T B^2 t_{co})-1$.

The synchrotron radiation is taken to be monochromatic, and the corresponding frequency of an electron with Lorentz factor $\gamma_e$ is $\nu_e = (3/2)\gamma_e^2\nu_L$, where $\nu_L=q_e B/(2\pi m_e c)$ is the Larmor frequency, and $q_e$ is the electron charge. The critical frequencies are $\nu_m = 3(1+z)^{-1}\gamma\gamma_m^2 q_e B/(2\pi m_e c)$ and $\nu_c= 3(1+z)^{-1}\gamma\gamma_c^2 q_e B/(2\pi m_e c)$, in the observer's frame respectively, where $\gamma$ is the bulk Lorentz factor of the emitted region. Before the reverse shock crosses the shell, $\nu_{c,2}=\nu_{c,3}$ is satisfied for the two regions having the same energy density $e$, Lorentz factor $\gamma$, and the same comoving time $t_{co}$.

The synchrotron self-absorption effect should not be ignored, especially at low frequencies, where the emission is modified enormously for the large optical depth. Wu et~al. (2003) have given the SSA coefficient and the corresponding spectral indices. We here quote the results of Wu et~al. (2003) and derive the SSA frequency for all six cases in the following.

The initial distribution of shock-accelerated electrons is
\begin{equation}
  N(\gamma_e)= N_{\gamma}\gamma_e^{-p} \phantom{-0o0-} \gamma_m < \gamma_e < \gamma_{\max}.
\end{equation}
Taking into account the synchrotron radiation energy losses, the power-law distribution of electrons is divided into two segments (Sari et~al. 1998), i.e., 
\begin{equation}
  N(\gamma_e)=N_\gamma \left\{
  \begin{array}{ll}
    \gamma_e ^{-2} &  \gamma_c < \gamma_e < \gamma_m\\
    \gamma_e^{-(p+1)} & \gamma_e > \gamma_m ,
  \end{array}
  \right.
\end{equation}
for the fast-cooling case ($\gamma_c < \gamma_m$), and
\begin{equation}
  N(\gamma_e)=N_\gamma \left\{
  \begin{array}{ll}
    \gamma_e ^{-p} & \gamma_m < \gamma_e < \gamma_c\\
    \gamma_e^{-(p+1)} & \gamma_e > \gamma_c ,
  \end{array}
  \right.
\end{equation}
for the slow-cooling case ($\gamma_c > \gamma_m$).

The self-absorption coefficients in different frequency ranges are
\begin{equation}
  \label{eq:A_k_nu}
  k_{\nu}=\frac{q_e}{B}N_\gamma \left\{
  \begin{array}{ll}
    c_1 \gamma_1^{-(p+4)} \left( \frac{\nu}{\nu_1} \right) ^{-5/3} & \nu \ll \nu_1 \\
    c_2 \gamma_1^{-(p+4)} \left( \frac{\nu}{\nu_1} \right) ^{-(p+4)/2} & \nu_1 \ll \nu \ll \nu_2, \\
    c_3 \gamma_2^{-(p+4)} \left( \frac{\nu}{\nu_2} \right) ^{-5/2} e ^{-\nu/\nu_2} & \nu \gg \nu_2 
  \end{array}
  \right.
\end{equation}
where $c_1=\frac{32 \pi^2}{9\times 2^{1/3}\Gamma(1/3)} \frac{p+2}{p+2/3}$, $c_2=\frac{2\sqrt{3}\pi}{9}2^{p/2} (p+\frac{10}{3})\Gamma(\frac{3p+2}{12}) \Gamma(\frac{3p+10}{12})$, $c_3=\frac{2\sqrt{6}\pi^{3/2}}{9}(p+2)$, $\nu_1$ and $\nu_2$ are the typical synchrotron frequencies of electrons of Lorentz factor $\gamma_1$ and $\gamma_2$, and $\Gamma(x)$ is the Gamma function (Wu et~al. 2003).

Electrons in both segments contribute to the SSA. For simplicity, the less important segment is neglected. In general, the absorption coefficient is dominated by the electrons between $\gamma_c$ and $\gamma_m$,  for the frequency less than $\max(\nu_c, \nu_m)$; and dominated by the electrons with Lorentz factor greater than $\max(\gamma_c, \gamma_m)$, for the frequency larger than $\max(\nu_c, \nu_m)$. So, the third expression in equation (\ref{eq:A_k_nu}) is always unimportant and can be neglected. We can obtain analytical expressions for the SSA frequency $\nu_a$ by taking $k_\nu L = \tau_0$,
\begin{equation}
  \nu_a = \left\{
  \begin{array}{lll}
    \left[ c_1 \frac{q_e}{B} N_\gamma \gamma_1^{-(p_1+4)} \nu_1^{5/3} \frac{L}{\tau_0} \right]^{\frac{3}{5}} & \nu_a \ll \nu_1\\
    \left[ c_2 \frac{q_e}{B} N_\gamma \gamma_1^{-(p_1+4)} \nu_1^{(p_1+4)/2} \frac{L}{\tau_0} \right]^{\frac{2}{p_1+4}} & \nu_1 \ll \nu_a \ll \nu_2\\
    \left[ c_2 \frac{q_e}{B} N_\gamma \gamma_2^{-(p_1+4)} \nu_2^{(p_2+4)/2} \frac{L}{\tau_0} \right]^{\frac{2}{p_2+4}} & \nu_a \gg \nu_2 ,
  \end{array}
  \right.
  \label{eq:A_nu_a}
\end{equation}
 where $L=N_e/(4\pi r^2n)$ is the comoving width of the emission region, $\tau_0$ can be defined equal to $0.35$ (Frail et~al. 2000), $\gamma_1 = \min(\gamma_c, \gamma_m)$, $\gamma_2 = \max(\gamma_c, \gamma_m)$, $p_1$ is the power-law index of the electron distribution between $\gamma_1$ and $\gamma_2$ ($p_1 = p$ for slowing cooling, $p_1 = 2$ for fast cooling), and $p_2=p+1$ is the index of the electron distribution with Lorentz factor greater than $\gamma_2$.

Because the peak spectral power $P_{\nu,\max}\simeq (1+z)\gamma m_e c^2 \sigma_T B/(3q_e)$ in the observer's frame is independent of $\gamma_e$, the peak observed flux density can be given by $F_{\nu,\max}=N_e P_{\nu,\max}/(4\pi D^2)$ at the frequency $\min(\nu_c,\nu_m)$, where $D$ is the luminosity distance of the gamma-ray burst.

\subsection{Relativistic Reverse Shock Case}\label{sec:relat-reverse-shock-1}

Using the above expressions, we obtain the typical frequencies and the peak flux density in the shocked shell and the shocked wind for the RRS case,
{\setlength\arraycolsep{2 pt}
\begin{eqnarray}
  \nu_{m,3} &\simeq& 5.9 \times 10^{15}(1+z)^{-1} \bar\epsilon_{e}^2 \epsilon_{B,-1}^{1/2} E_{53}^{-1/2} \eta_{2.5}^2 A_{*,-1} \Delta_{0,12}^{-1/2} \left(\frac{t_{\oplus}}{T}\right)^{-1} ~\mbox{Hz},\\
  \nu_{m,2} &\simeq& 1.0 \times 10^{19} (1+z)^{-1} \bar\epsilon_{e}^2 \epsilon_{B,-1}^{1/2} E_{53}^{1/2} \Delta_{0,12}^{-3/2} \left(\frac{t_{\oplus}}{T}\right)^{-1} ~\mbox{Hz}\\
  \nu_{c,2} &=& \nu_{c,3} \simeq 1.5 \times 10^{12}(1+z)^{-1} \epsilon_{B,-1}^{-3/2} E_{53}^{1/2}\Delta_{0,12}^{1/2} A_{*,-1}^{-2} \left.\frac{t_{\oplus}}{T}\right. ~\mbox{Hz} ,\\
  F_{\nu,\max,3} &\simeq& 95.3  (1+z)\epsilon_{B,-1}^{1/2} E_{53} \eta_{2.5}^{-1} A_{*,-1}^{1/2} \Delta_{0,12}^{-1}D_{28}^{-2} ~\mbox{Jy},\\
  F_{\nu,\max,2} &\simeq& 1.6  (1+z)\epsilon_{B,-1}^{1/2} E_{53}^{1/2} A_{*,-1} \Delta_{0,12}^{-1/2} D_{28} ^{-2} ~\mbox{Jy}
\end{eqnarray}}
where $\bar\epsilon_{e} \equiv \epsilon_{e,-0.5}\cdot 3(p-2)/(p-1)$. Note that $F_{\nu,\max,3}>F_{\nu,\max,2}$, i.e. region 3 dominates the emission for the early afterglow, mainly because the number of electrons in region 3 is much larger than that in region2.

We give the scaling-law of the SSA frequency in region 3,
\begin{equation}
  \nu_{a,3} \simeq \left\{
  \begin{array}{ll}
    1.1 \times 10^{16}(1+z)^{-1} \epsilon_{B,-1}^{6/5} E_{53}^{-1/10} \eta_{2.5}^{-3/5} A_{*,-1}^{19/10}\Delta_{0,12}^{-19/10} \left(\frac{t_{\oplus}}{T}\right)^{-2} ~\mbox{Hz} & \nu_a < \nu_c < \nu_m \\
    2.2 \times 10^{14}(1+z)^{-1} E_{53}^{1/6} \eta_{2.5}^{-1/3} A_{*,-1}^{1/6}\Delta_{0,12}^{-5/6} \left(\frac{t_{\oplus}}{T}\right)^{-2/3} ~\mbox{Hz} & \nu_c < \nu_a < \nu_m \\
    4.7 \times 10^{14}(1+z)^{-1} \bar \epsilon_{e}^{2/5} \epsilon_{B,-1}^{1/10} E_{53}^{1/30} \eta_{2.5}^{2/15} A_{*,-1}^{1/3}\Delta_{0,12}^{-23/30} \left(\frac{t_{\oplus}}{T}\right)^{-11/15} ~\mbox{Hz} & \nu_c < \nu_m < \nu_a,\\

    2.2 \times 10^{14} (1+z)^{-1}\bar \epsilon_{e}^{-1} \epsilon_{B,-1}^{1/5} E_{53}^{2/5} \eta_{2.5}^{-8/5} A_{*,-1}^{2/5}\Delta_{0,12}^{-7/5} \left(\frac{t_{\oplus}}{T}\right)^{-1}  ~\mbox{Hz} & \nu_a < \nu_m < \nu_c\\
    1.1  \times 10^{15}(1+z)^{-1} \bar \epsilon_{e}^{6/13} \epsilon_{B,-1}^{9/26} E_{53}^{-1/26} \eta_{2.5}^{2/13} A_{*,-1}^{9/13}\Delta_{0,12}^{-25/26} \left(\frac{t_{\oplus}}{T}\right)^{-1} ~\mbox{Hz} & \nu_m < \nu_a < \nu_c\\
    5.2  \times 10^{14}(1+z)^{-1} \bar \epsilon_{e}^{2/5} \epsilon_{B,-1}^{1/10} E_{53}^{1/30} \eta_{2.5}^{2/15} A_{*,-1}^{1/3}\Delta_{0,12}^{-23/30} \left(\frac{t_{\oplus}}{T}\right)^{-11/15} ~\mbox{Hz} & \nu_m < \nu_c < \nu_a,
  \end{array}
  \right.
\end{equation}
Here and in the following expressions for $\nu_a$, we take $p=2.5$. If more than one expressions above satisfy the followed restriction, the largest $\nu_a$ is the true value.

In region 2,
\begin{equation}
  \nu_{a,2} \simeq \left\{
  \begin{array}{ll}
    9.6 \times 10^{14}  (1+z)^{-1}\epsilon_{B,-1}^{6/5} E_{53}^{-2/5} A_{*,-1}^{11/5}\Delta_{0,12}^{-8/5} \left(\frac{t_{\oplus}}{T}\right)^{-2}  ~\mbox{Hz} & \nu_a < \nu_c < \nu_m \\
    5.7 \times 10^{13}  (1+z)^{-1} A_{*,-1}^{1/3}\Delta_{0,12}^{-2/3} \left(\frac{t_{\oplus}}{T}\right)^{-2/3}  ~\mbox{Hz} & \nu_c < \nu_a < \nu_m \\
    7.0 \times 10^{14}  (1+z)^{-1} \bar\epsilon_{e}^{2/5} \epsilon_{B,-1}^{1/10} E_{53}^{1/10} A_{*,-1}^{4/15}\Delta_{0,12}^{-5/6} \left(\frac{t_{\oplus}}{T}\right)^{-11/15}  ~\mbox{Hz} & \nu_c < \nu_m < \nu_a,\\
    
    4.5 \times 10^{11}  (1+z)^{-1} \bar\epsilon_{e}^{-1} \epsilon_{B,-1}^{1/5} E_{53}^{-2/5} A_{*,-1}^{6/5}\Delta_{0,12}^{-3/5} \left(\frac{t_{\oplus}}{T}\right)^{-1}  ~\mbox{Hz} & \nu_a < \nu_m < \nu_c\\
    1.9 \times 10^{15}  (1+z)^{-1} \bar\epsilon_{e}^{6/13} \epsilon_{B,-1}^{9/26} E_{53}^{1/26} A_{*,-1}^{8/13}\Delta_{0,12}^{-27/26} \left(\frac{t_{\oplus}}{T}\right)^{-1} ~\mbox{Hz}  & \nu_m < \nu_a < \nu_c\\
    7.8 \times 10^{14} (1+z)^{-1} \bar \epsilon_{e}^{2/5} \epsilon_{B,-1}^{1/10} E_{53}^{1/10} A_{*,-1}^{4/15}\Delta_{0,12}^{-5/6} \left(\frac{t_{\oplus}}{T}\right)^{-11/15} ~\mbox{Hz}  & \nu_m < \nu_c < \nu_a.
  \end{array}
  \right.
\end{equation}

After the reverse shock crosses the shell ($t_{\oplus} > T$), the behavior of both shocked regions can be described by the BM self-similar solution. The power-law indices of emission variables with time are given in table \ref{tab:var_wind}. Another frequency $\nu_{cut}$ should be introduced here (Kobayashi 2000) to substitute $\nu_c$ for no fresh electrons. $\nu_{cut}$ has the same time profile as $\nu_m$. If $\nu_m > \nu_{cut}$, $\nu_m$ comes down to $\nu_{cut}$ for the synchrotron cooling. And if $\nu_a > \nu_{cut}$, $\nu_a$ comes down to $\nu_{cut}$ too for no electrons distributed greater than corresponding $\gamma_{cut}$. These are all represented in columns labeled \textcircled{\scriptsize6} (for NRS case) and \textcircled{\scriptsize8} in Table~\ref{tab:var_wind}.

The scaling-law indices of flux densities with time are sophisticated, as they vary with time when any two of $\nu, \nu_a, \nu_c (\mbox{or } \nu_{cut}), \nu_m $ cross each other, where $\nu$ is the observed frequency. These indices are given in Table~\ref{tab:flux_wind}. For the case $\nu > \nu_{cut}$, the flux density decreases exponentially with observed frequency $\nu$, then we take it to be zero, which is denoted by a short horizontal line in Table~\ref{tab:flux_wind}. The numerical results will be given in \S \ref{sec:numerical-results}.

For the typical parameters, the order of the frequencies at $t_{\oplus} = T$ is $\nu_m > \nu >\nu_a > \nu_c$ for both region 3 and region 2, if the considered frequency is $\nu = 4.55\times 10^{14} $Hz. The flux density from the shocked shell and shocked environment are
\begin{equation}
  F_{\nu,3} \simeq 5.4 (1+z)^{1/2} \epsilon_{B,-1}^{-1/4} E_{53}^{5/4} \eta_{2.5}^{-1} A_{*,-1}^{-1/2} \Delta_{0,12}^{-3/4} D_{28}^{-2} ~\mbox{Jy}.
  \label{eq:flux3f23_RRS}
\end{equation}
\begin{equation}
  F_{\nu,2} \simeq 0.1 (1+z)^{1/2} \epsilon_{B,-1}^{-1/4} E_{53}^{3/4} \Delta_{0,12}^{-1/4} D_{28}^{-2}  ~\mbox{Jy}.
  \label{eq:flux2f23_RRS}
\end{equation}

\subsection{Newtonian Reverse Shock Case}\label{sec:Newton-reverse-shock-1}

For NRS case, before the reverse shock crosses the shell ($t_{\oplus}< T'$),
{\setlength\arraycolsep{2 pt}
\begin{eqnarray}
  \nu_{m,3} &\simeq& 4.1 \times 10^{12} (1+z)^{-1} \bar\epsilon_{e}^2 \epsilon_{B,-1}^{1/2} E_{53}^{-1} \eta_{1.5}^{4} A_{*,-1}^{3/2} \left.\frac{t_{\oplus}}{T'}\right. ~\mbox{Hz}\\
  \nu_{m,2} &\simeq& 1.3  \times 10^{16} (1+z)^{-1}  \bar\epsilon_{e}^2 \epsilon_{B,-1}^{1/2} E_{53}^{-1} \eta_{1.5}^6 A_{*,-1}^{3/2} \left(\frac{t_{\oplus}}{T'}\right)^{-1} ~\mbox{Hz}\\
  \nu_{c,3} &=& \nu_{c,2}  \simeq 2.7 \times 10^{13} (1+z)^{-1} \epsilon_{B,-1}^{-3/2} E_{53} \eta_{1.5}^{-2} A_{*,-1}^{-5/2} \left.\frac{t_{\oplus}}{T'}\right. ~\mbox{Hz}\\
  F_{\nu,\max,3} &\simeq& 7.6 (1+z) \epsilon_{B,-1}^{1/2} \eta_{1.5}^{3} A_{*,-1}^{3/2} D_{28}^{-2} \left(\frac{t_{\oplus}}{T'}\right)^{-1/2} ~\mbox{Jy}\\
  F_{\nu,\max,2} &\simeq& 0.2 (1+z) \epsilon_{B,-1}^{1/2} \eta_{1.5}^{2} A_{*,-1}^{3/2} D_{28}^{-2} ~\mbox{Jy}
\end{eqnarray}}

The SSA frequency in region 3,
\begin{equation}
  \nu_{a,3} \simeq \left\{
  \begin{array}{ll}
    2.2 \times 10^{12} (1+z)^{-1} \epsilon_{B,-1}^{6/5} E_{53}^{-2} \eta_{1.5}^{7} A_{*,-1}^{19/5} \left(\frac{t_{\oplus}}{T'}\right)^{-23/10} ~\mbox{Hz}  & \nu_a < \nu_c < \nu_m \\
    7.2 \times 10^{12} (1+z)^{-1} E_{53}^{-2/3} \eta_{1.5}^{3} A_{*,-1} \left(\frac{t_{\oplus}}{T'}\right)^{-5/6} ~\mbox{Hz}  & \nu_c < \nu_a < \nu_m \\
    7.1 \times 10^{12} (1+z)^{-1}  \bar\epsilon_{e}^{2/5} \epsilon_{B,-1}^{1/10} E_{53}^{-11/15} \eta_{1.5}^{16/5} A_{*,-1}^{11/10} \left(\frac{t_{\oplus}}{T'}\right)^{-7/15} ~\mbox{Hz}  & \nu_c < \nu_m < \nu_a,\\
    
    7.3 \times 10^{12} (1+z)^{-1} \bar \epsilon_{e}^{-1} \epsilon_{B,-1}^{1/5} E_{53}^{-1} \eta_{1.5}^{4} A_{*,-1}^{9/5} \left(\frac{t_{\oplus}}{T'}\right)^{-23/10} ~\mbox{Hz}  & \nu_a < \nu_m < \nu_c\\
    5.8 \times 10^{12} (1+z)^{-1} \bar \epsilon_{e}^{6/13} \epsilon_{B,-1}^{9/26} E_{53}^{-1} \eta_{1.5}^{4} A_{*,-1}^{43/26} \left(\frac{t_{\oplus}}{T'}\right)^{-9/13} ~\mbox{Hz}  & \nu_m < \nu_a < \nu_c\\
    7.9 \times 10^{12}  (1+z)^{-1} \bar\epsilon_{e}^{2/5} \epsilon_{B,-1}^{1/10} E_{53}^{-11/15} \eta_{1.5}^{16/5} A_{*,-1}^{11/10} \left(\frac{t_{\oplus}}{T'}\right)^{-7/15} ~\mbox{Hz}  & \nu_m < \nu_c < \nu_a,
  \end{array}
  \right.
\end{equation}
and in region 2,
\begin{equation}
  \nu_{a,2} \simeq \left\{
  \begin{array}{ll}
    2.8 \times 10^{11} (1+z)^{-1} \epsilon_{B,-1}^{6/5} E_{53}^{-2} \eta_{1.5}^{32/5} A_{*,-1}^{19/5} \left(\frac{t_{\oplus}}{T'}\right)^{-2} ~\mbox{Hz}  & \nu_a < \nu_c < \nu_m \\
    2.3 \times 10^{12} (1+z)^{-1} E_{53}^{-2/3} \eta_{1.5}^{8/3} A_{*,-1} \left(\frac{t_{\oplus}}{T'}\right)^{-2/3} ~\mbox{Hz}  & \nu_c < \nu_a < \nu_m \\
    1.4 \times 10^{13} (1+z)^{-1} \bar \epsilon_{e}^{2/5} \epsilon_{B,-1}^{1/10} E_{53}^{-11/15} \eta_{1.5}^{10/3} A_{*,-1}^{11/10} \left(\frac{t_{\oplus}}{T'}\right)^{-11/15} ~\mbox{Hz}  & \nu_c < \nu_m < \nu_a,\\
    
    1.7 \times 10^{10}  (1+z)^{-1} \bar\epsilon_{e}^{-1} \epsilon_{B,-1}^{1/5} E_{53}^{-1} \eta_{1.5}^{12/5} A_{*,-1}^{9/5} \left(\frac{t_{\oplus}}{T'}\right)^{-1} ~\mbox{Hz}  & \nu_a < \nu_m < \nu_c\\
    1.3 \times 10^{13}  (1+z)^{-1} \bar\epsilon_{e}^{6/13} \epsilon_{B,-1}^{9/26} E_{53}^{-1} \eta_{1.5}^{54/13} A_{*,-1}^{43/26} \left(\frac{t_{\oplus}}{T'}\right)^{-1}  ~\mbox{Hz}  & \nu_m < \nu_a < \nu_c\\
    1.6 \times 10^{13}  (1+z)^{-1} \bar\epsilon_{e}^{2/5} \epsilon_{B,-1}^{1/10} E_{53}^{-11/15} \eta_{1.5}^{10/3} A_{*,-1}^{11/10} \left(\frac{t_{\oplus}}{T'}\right)^{-11/15}  ~\mbox{Hz}  & \nu_m < \nu_c < \nu_a,
  \end{array}
  \right.
\end{equation}

After the reverse shock crosses the shell ($t_{\oplus}>T'$), the temporal indices of the typical frequencies and the observed flux density are also given in tables \ref{tab:var_wind} and \ref{tab:flux_wind}.

The frequency relations at time $t_{\oplus} = T'$ are $\nu > \nu_c >\nu_a > \nu_m$ for region 3 and $\nu_m > \nu >\nu_c > \nu_a$ for region 2, if $\nu = 4.55\times 10^{14} $Hz. The corresponding optical flux density from region 3 and 2 are
\begin{equation}
  F_{\nu,3} \simeq 65 (1+z)^{-1/4} \bar \epsilon_{e}^{3/2} \epsilon_{B,-1}^{1/8} E_{53}^{-1/4} \eta_{1.5}^{5} A_{*,-1}^{11/8} D_{28}^{-2}  ~\mbox{mJy},
  \label{eq:flux3f33_NRS}
\end{equation}
\begin{equation}
  F_{\nu,2} \simeq 63 (1+z)^{1/2} \epsilon_{B,-1}^{-1/4} E_{53}^{1/2} \eta_{1.5} A_{*,-1}^{1/4} D_{28}^{-2}  ~\mbox{mJy}.
  \label{eq:flux2f23_NRS}
\end{equation}

\begin{table}
\caption{The temporal indices for the evolution of $\nu_m$, $\nu_c$, $\nu_a$ and $F_{\nu,\max}$. The notations denote respectively: 
  \textcircled{\scriptsize1} early, NRS, region 2;
  \textcircled{\scriptsize2} early, NRS, region 3;
  \textcircled{\scriptsize3} early, RRS, region 2;
  \textcircled{\scriptsize4} early, RRS, region 3;
  \textcircled{\scriptsize5} lately, NRS, region 2;
  \textcircled{\scriptsize6} lately, NRS, region 3;
  \textcircled{\scriptsize7} lately, RRS, region 2;
  \textcircled{\scriptsize8} lately, RRS, region 3.
$\nu_c$ is actually $\nu_{cut}$ at the columns \textcircled{\scriptsize6} \& \textcircled{\scriptsize8}.}
\label{tab:var_wind}
\begin{tabular}{ccccccc} \hline
  variable & $t < t_{\Delta}$ &  &  & $t > t_{\Delta}$ &  & notation\\
  &  \textcircled{\scriptsize1} \textcircled{\scriptsize3} \textcircled{\scriptsize4} &  \textcircled{\scriptsize2} &  \textcircled{\scriptsize5} \textcircled{\scriptsize7} &  \textcircled{\scriptsize6} &  \textcircled{\scriptsize8} & \\ \hline 
  $\nu_m$ & $- 1$ & $1$ & $- \frac{3}{2}$ & $- \frac{15 \hspace{0.25em} g +
  24}{14 \hspace{0.25em} g + 7}$ & $- \frac{15}{8}$ & \\
  $\nu_c$ & $1$ & $1$ & $\frac{1}{2}$ & $- \frac{15 \hspace{0.25em} g + 24}{14
  \hspace{0.25em} g + 7}$ & $- \frac{15}{8}$ & \\
  & $- 2$ & $- \frac{23}{10}$ & $- \frac{8}{5}$ & $- \frac{33 \hspace{0.25em}
  g + 36}{70 \hspace{0.25em} g + 35}$ & $- \frac{3}{5}$ & $\nu_a \ll \nu_c \ll
  \nu_m$\\
  $\nu_a$ & $- \frac{2}{3}$ & $- \frac{5}{6}$ & $- \frac{2}{3}$ & $- \frac{15
  \hspace{0.25em} g + 24}{14 \hspace{0.25em} g + 7}$ & $- \frac{15}{8}$ &
  $\nu_c \ll \nu_a \ll \nu_m$\\
  (fast cooling) & $- \frac{p + 3}{p + 5}$ & $\frac{p - 6}{p + 5}$ & $-
  \frac{3 \hspace{0.25em} p + 5}{2 \hspace{0.25em} p + 10}$ & $- \frac{15
  \hspace{0.25em} g + 24}{14 \hspace{0.25em} g + 7}$ & $- \frac{15}{8}$ &
  $\nu_c \ll \nu_m \ll \nu_a$\\
  & $- 1$ & $- \frac{23}{10}$ & $- \frac{3}{5}$ & $- \frac{33 \hspace{0.25em}
  g + 36}{70 \hspace{0.25em} g + 35}$ & $- \frac{3}{5}$ & $\nu_a \ll \nu_m \ll
  \nu_c$\\
  $\nu_a$ & $- 1$ & $\frac{p - 7}{p + 4}$ & $- \frac{3 \hspace{0.25em} p +
  6}{2 \hspace{0.25em} p + 8}$ & $- \frac{\left( 15 \hspace{0.25em} g + 24
  \right) \hspace{0.25em} p + 32 \hspace{0.25em} g + 40}{\left( 14
  \hspace{0.25em} g + 7 \right) \hspace{0.25em} p + 56 \hspace{0.25em} g +
  28}$ & $- \frac{15 \hspace{0.25em} p + 26}{8 \hspace{0.25em} p + 32}$ &
  $\nu_m \ll \nu_a \ll \nu_c$\\
  (slow cooling) & $- \frac{p + 3}{p + 5}$ & $\frac{p - 6}{p + 5}$ & $-
  \frac{3 \hspace{0.25em} p + 5}{2 \hspace{0.25em} p + 10}$ & $- \frac{15
  \hspace{0.25em} g + 24}{14 \hspace{0.25em} g + 7}$ & $- \frac{15}{8}$ &
  $\nu_m \ll \nu_c \ll \nu_a$\\
  $F_{\nu, \max}$ & $0$ & $- \frac{1}{2}$ & $- \frac{1}{2}$ & $- \frac{11
  \hspace{0.25em} g + 12}{14 \hspace{0.25em} g + 7}$ & $- \frac{9}{8}$ & \\ \hline
\end{tabular}
\end{table}

\begin{table}
\caption{The temporal indices for the evolution of flux density ($F_{\nu} \propto t_\oplus^\alpha$). The notations are the same as in table \ref{tab:var_wind}. The short horizontal lines indicate the radiation vanishes at those cases.}
\label{tab:flux_wind}
\begin{tabular}{cccccc} \hline
  & $t < t_{\Delta}$ &  &  & $t > t_{\Delta}$ & \\
  case & \textcircled{\scriptsize1}\textcircled{\scriptsize3}\textcircled{\scriptsize4} & \textcircled{\scriptsize2} & \textcircled{\scriptsize5}\textcircled{\scriptsize7} & \textcircled{\scriptsize6} & \textcircled{\scriptsize8}\\ \hline
  $\nu < \nu_a < \nu_c < \nu_m$ & $3$ & $3$ & $2$ & $\frac{5 \hspace{0.25em} g
  + 8}{14 \hspace{0.25em} g + 7}$ & $\frac{1}{2}$\\
  $\nu_a < \nu_{} < \nu_c < \nu_m$ & $- \frac{1}{3}$ & $- \frac{5}{6}$ & $-
  \frac{2}{3}$ & $- \frac{6 \hspace{0.25em} g + 4}{14 \hspace{0.25em} g + 7}$
  & $- \frac{1}{2}$\\
  $\nu_a < \nu_c < \nu < \nu_m$ & $\frac{1}{2}$ & $0$ & $- \frac{1}{4}$ & -- &
  --\\
  $\nu_a < \nu_c < \nu_m < \nu$ & $- \frac{p - 2}{2}$ & $\frac{p - 1}{2}$ & $-
  \frac{3 \hspace{0.25em} p - 2}{4}$ & -- & --\\
  $\nu < \nu_c < \nu_a < \nu_m$ & $3$ & $3$ & $2$ & $\frac{19 \hspace{0.25em}
  g + 36}{14 \hspace{0.25em} g + 7}$ & $\frac{21}{8}$\\
  $\nu_c < \nu_{} < \nu_a < \nu_m$ & $\frac{5}{2}$ & $\frac{5}{2}$ &
  $\frac{7}{4}$ & -- & --\\
  $\nu_c < \nu_a < \nu_{} < \nu_m$ & $\frac{1}{2}$ & $0$ & $- \frac{1}{4}$ &
  -- & --\\
  $\nu_c < \nu_a < \nu_m < \nu_{}$ & $- \frac{p - 2}{2}$ & $\frac{p - 1}{2}$ &
  $- \frac{3 \hspace{0.25em} p - 2}{4}$ & -- & --\\
  $\nu < \nu_c < \nu_m < \nu_a$ & $3$ & $3$ & $2$ & -- & --\\
  $\nu_c < \nu_{} < \nu_a$ & $\frac{5}{2}$ & $\frac{5}{2}$ & $\frac{7}{4}$ &
  -- & --\\
  $\nu_c < \nu_m < \nu_a < \nu_{}$ & $- \frac{p - 2}{2}$ & $\frac{p - 1}{2}$ &
  $- \frac{3 \hspace{0.25em} p - 2}{4}$ & -- & --\\
  $\nu < \nu_a < \nu_m < \nu_c$ & $2$ & $3$ & $1$ & $\frac{5 \hspace{0.25em} g
  + 8}{14 \hspace{0.25em} g + 7}$ & $\frac{1}{2}$\\
  $\nu_a < \nu < \nu_m < \nu_c$ & $\frac{1}{3}$ & $- \frac{5}{6}$ & $0$ & $-
  \frac{6 \hspace{0.25em} g + 4}{14 \hspace{0.25em} g + 7}$ & $-
  \frac{1}{2}$\\
  $\nu_a < \nu_m < \nu < \nu_c$ & $- \frac{p - 1}{2}$ & $\frac{p - 2}{2}$ & $-
  \frac{3 \hspace{0.25em} p - 1}{4}$ & $- \frac{\left( 15 \hspace{0.25em} g +
  24 \right) \hspace{0.25em} p + 7 \hspace{0.25em} g}{28 \hspace{0.25em} g +
  14}$ & $- \frac{15 \hspace{0.25em} p + 3}{16}$\\
  $\nu_a < \nu_m < \nu_c < \nu_{}$ & $- \frac{p - 2}{2}$ & $\frac{p - 1}{2}$ &
  $- \frac{3 \hspace{0.25em} p - 2}{4}$ & -- & --\\
  $\nu < \nu_m < \nu_a < \nu_c$ & $2$ & $3$ & $1$ & $\frac{5 \hspace{0.25em} g
  + 8}{14 \hspace{0.25em} g + 7}$ & $\frac{1}{2}$\\
  $\nu_m < \nu < \nu_a < \nu_c$ & $\frac{5}{2}$ & $\frac{5}{2}$ &
  $\frac{7}{4}$ & $\frac{25 \hspace{0.25em} g + 40}{28 \hspace{0.25em} g +
  14}$ & $\frac{23}{16}$\\
  $\nu_m < \nu_a < \nu < \nu_c$ & $- \frac{p - 1}{2}$ & $\frac{p - 2}{2}$ & $-
  \frac{3 \hspace{0.25em} p - 1}{4}$ & $- \frac{\left( 15 \hspace{0.25em} g +
  24 \right) \hspace{0.25em} p + 7 \hspace{0.25em} g}{28 \hspace{0.25em} g +
  14}$ & $- \frac{15 \hspace{0.25em} p + 3}{16}$\\
  $\nu_m < \nu_a < \nu_c < \nu$ & $- \frac{p - 2}{2}$ & $\frac{p - 1}{2}$ & $-
  \frac{3 \hspace{0.25em} p - 2}{4}$ & -- & --\\
  $\nu < \nu_m < \nu_c < \nu_a$ & $2$ & $3$ & $1$ & $\frac{19 \hspace{0.25em}
  g + 36}{14 \hspace{0.25em} g + 7}$ & $\frac{21}{8}$\\
  $\nu_m < \nu < \nu_a$ & $\frac{5}{2}$ & $\frac{5}{2}$ & $\frac{7}{4}$ &
  $\frac{53 \hspace{0.25em} g + 96}{28 \hspace{0.25em} g + 14}$ &
  $\frac{57}{16}$\\
  $\nu_m < \nu_c < \nu_a < \nu_{}$ & $- \frac{p - 2}{2}$ & $\frac{p - 1}{2}$ &
  $- \frac{3 \hspace{0.25em} p - 2}{4}$ & -- & --\\ \hline
\end{tabular}
\end{table}

\section{Numerical results}\label{sec:numerical-results}

The above analytical results can give approximate behaviors of variables as functions of time or frequency, but they are valid only in relativistic or Newtonian limits. In the mildly relativistic case, the analytical values deviate from the actual ones very much (Nakar 2004). For precise results, a numerical method should be engaged.

Combining equations (\ref{eq:jump1}), (\ref{eq:jump2}) and (\ref{eq:gamma_34}), and using the assumption of equalities between the Lorentz factors and pressures beside the surface of the contact discontinuity, one can obtain solutions of $\gamma_2$, $\gamma_3$, $\bar{\gamma_3}$, $e_2$, $n_2$, $n_3$ numerically. Before the reverse shock crosses the shell, the value of $\gamma_3$ should be solved from the following equation without approximation,
{\setlength\arraycolsep{2 pt}
\begin{eqnarray}
  (\gamma_3-1)(4\gamma_3+3) &=& \left[ \gamma_3 \gamma_4 \left( 1-\sqrt{1-\frac{1}{\gamma_3^2}}\sqrt{1-\frac{1}{\gamma_4^2}}\right) -1\right] \nonumber\\ 
   && \times \left[ 4\gamma_3 \gamma_4 \left( 1- \sqrt{1-\frac{1}{\gamma_3^2}}\sqrt{1-\frac{1}{\gamma_4^2}}\right) +3 \right]f,
\end{eqnarray}}
and then the other variables can be derived directly.

\begin{figure}
  \centering
  \includegraphics[angle=270,width=0.4\textwidth]{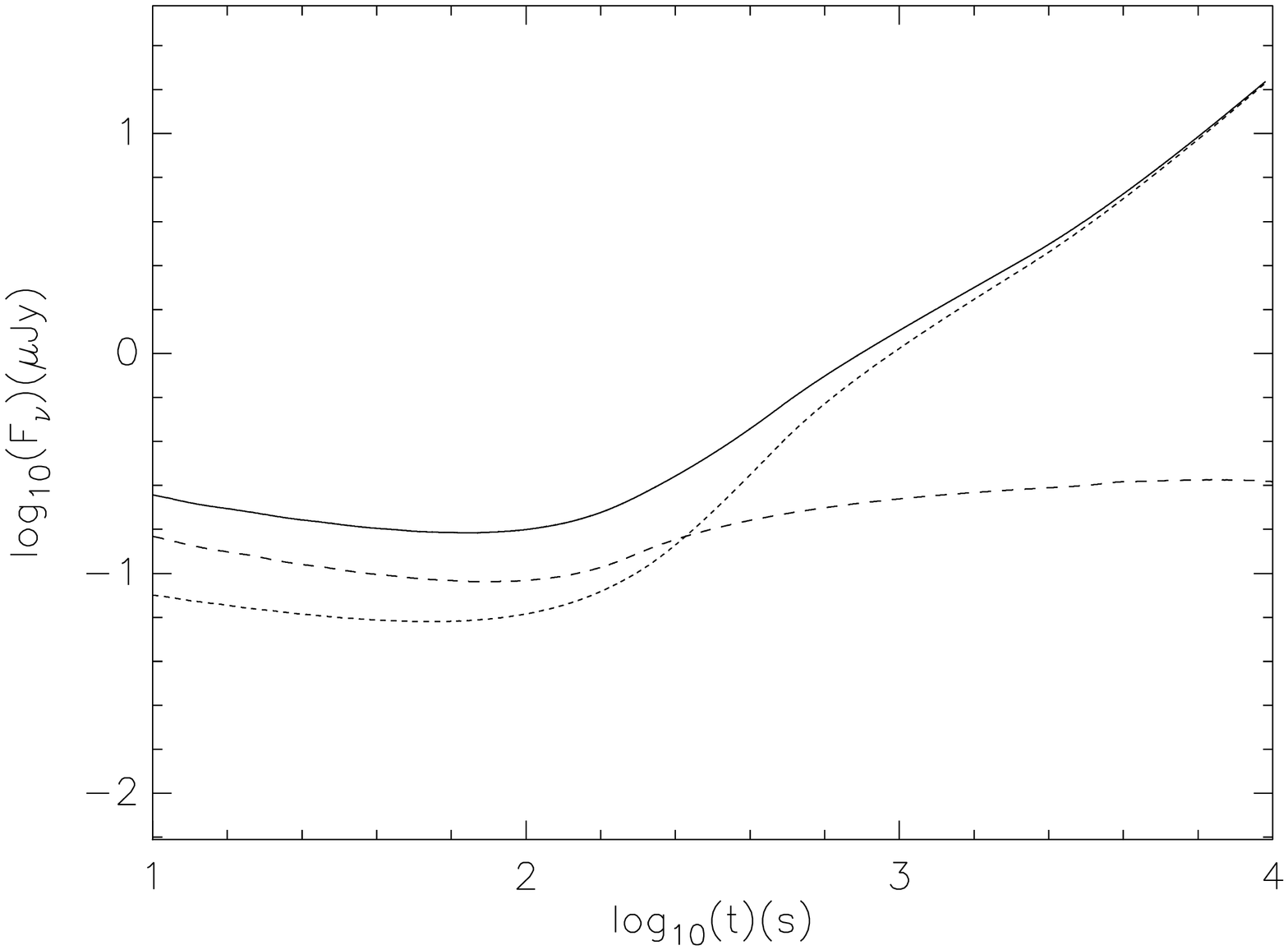}\\
  \includegraphics[angle=270,width=0.4\textwidth]{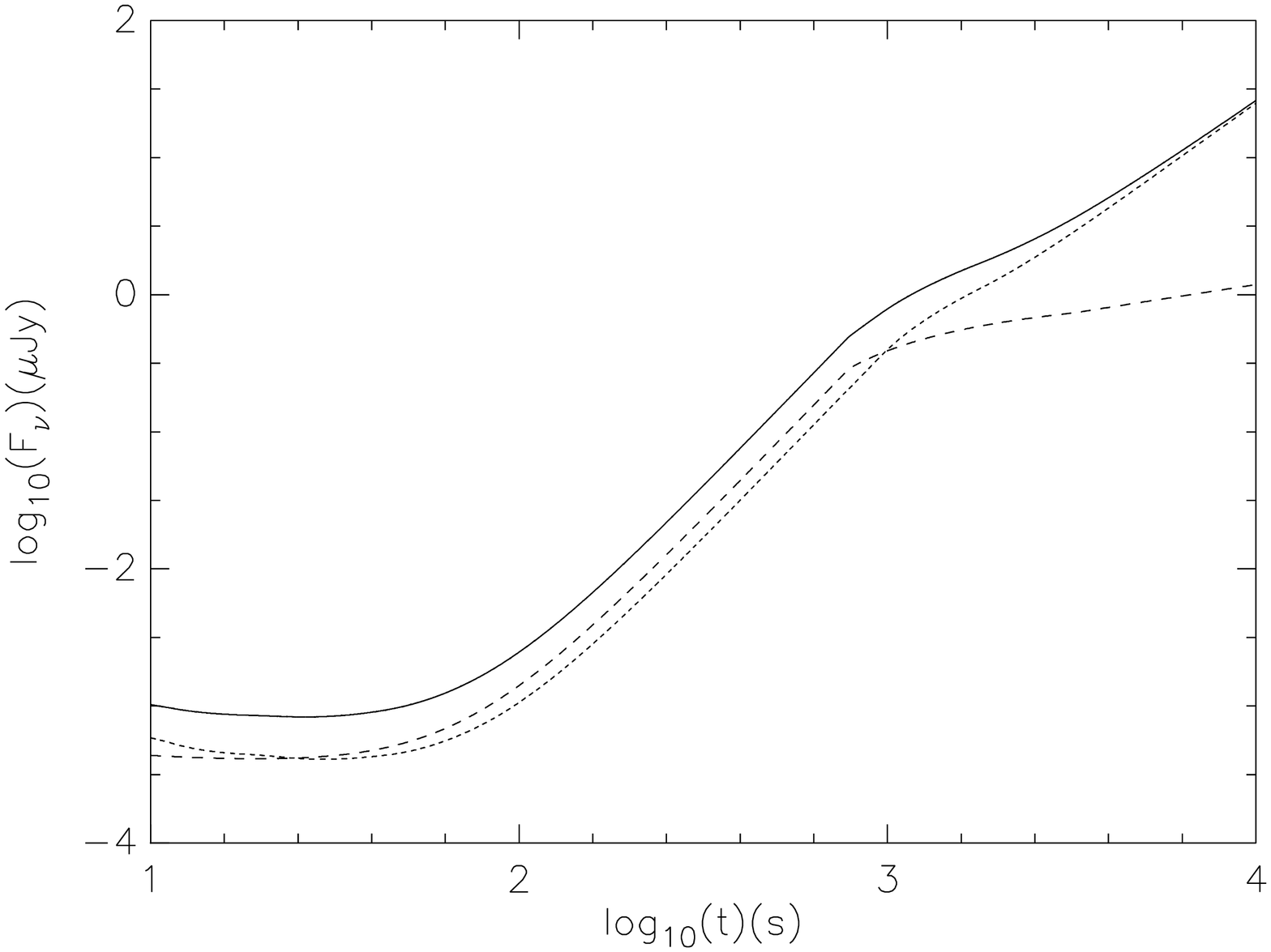}
  \caption{Flux density at $\nu=8.46$ GHz as function of time. Parameters are $\eta=300, E_0=1.0\times10^{52}~\mbox{erg}, A_{*}=0.1, \Delta_0=5.0\times10^{12}~\mbox{cm}, \epsilon_e=0.3, \epsilon_B=0.1, p=2.5$ for the upper panel (RRS case). Only $\eta=30$ is different for the lower panel (NRS case). The long dashed and short dashed lines represent the emission from region 3 and region 2, respectively, and the solid line is the total flux density from both regions.}
  \label{fig:f_radio}
\end{figure}

\begin{figure}
  \centering
  \includegraphics[angle=270,width=0.4\textwidth]{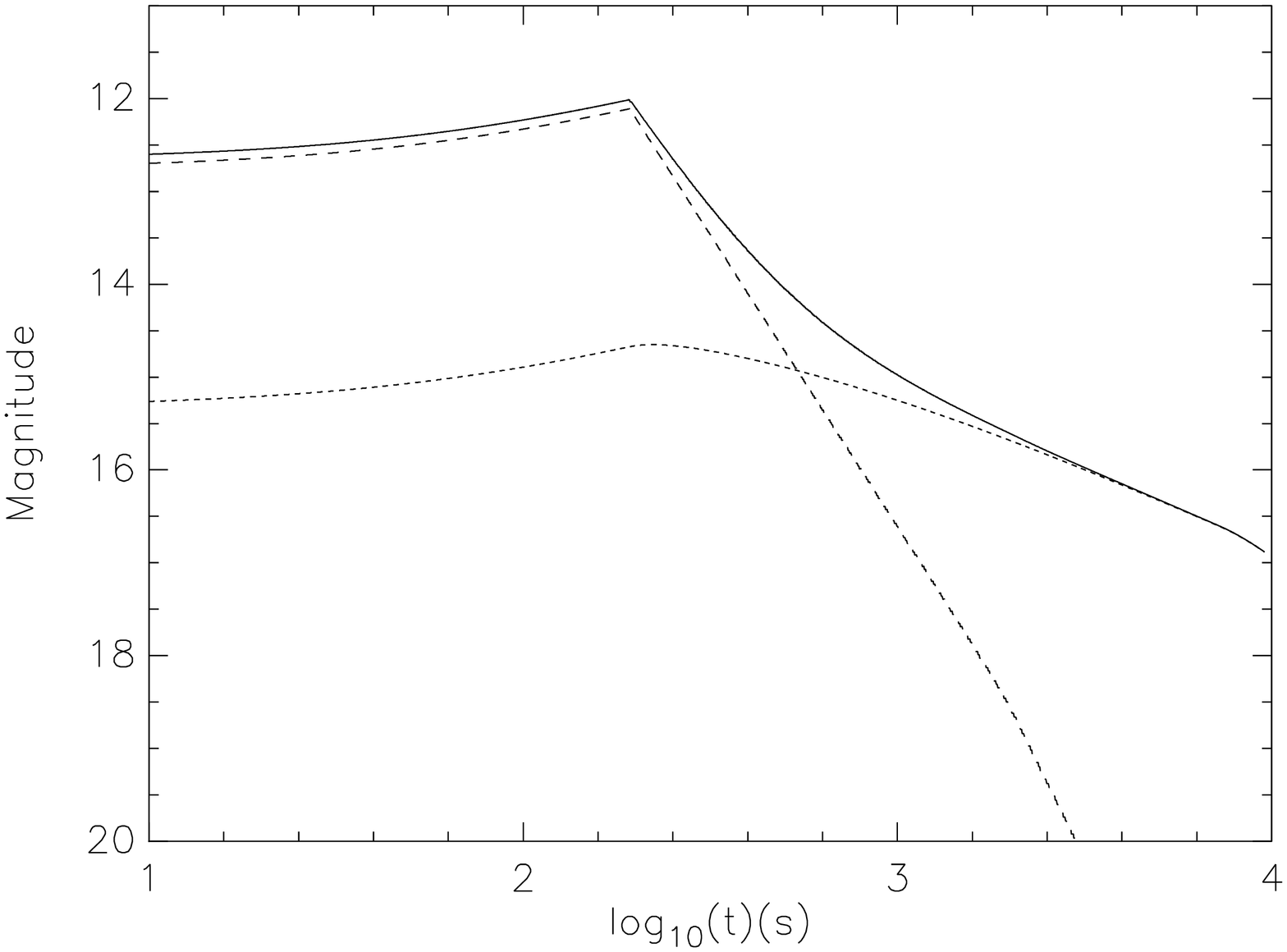}\\
  \includegraphics[angle=270,width=0.4\textwidth]{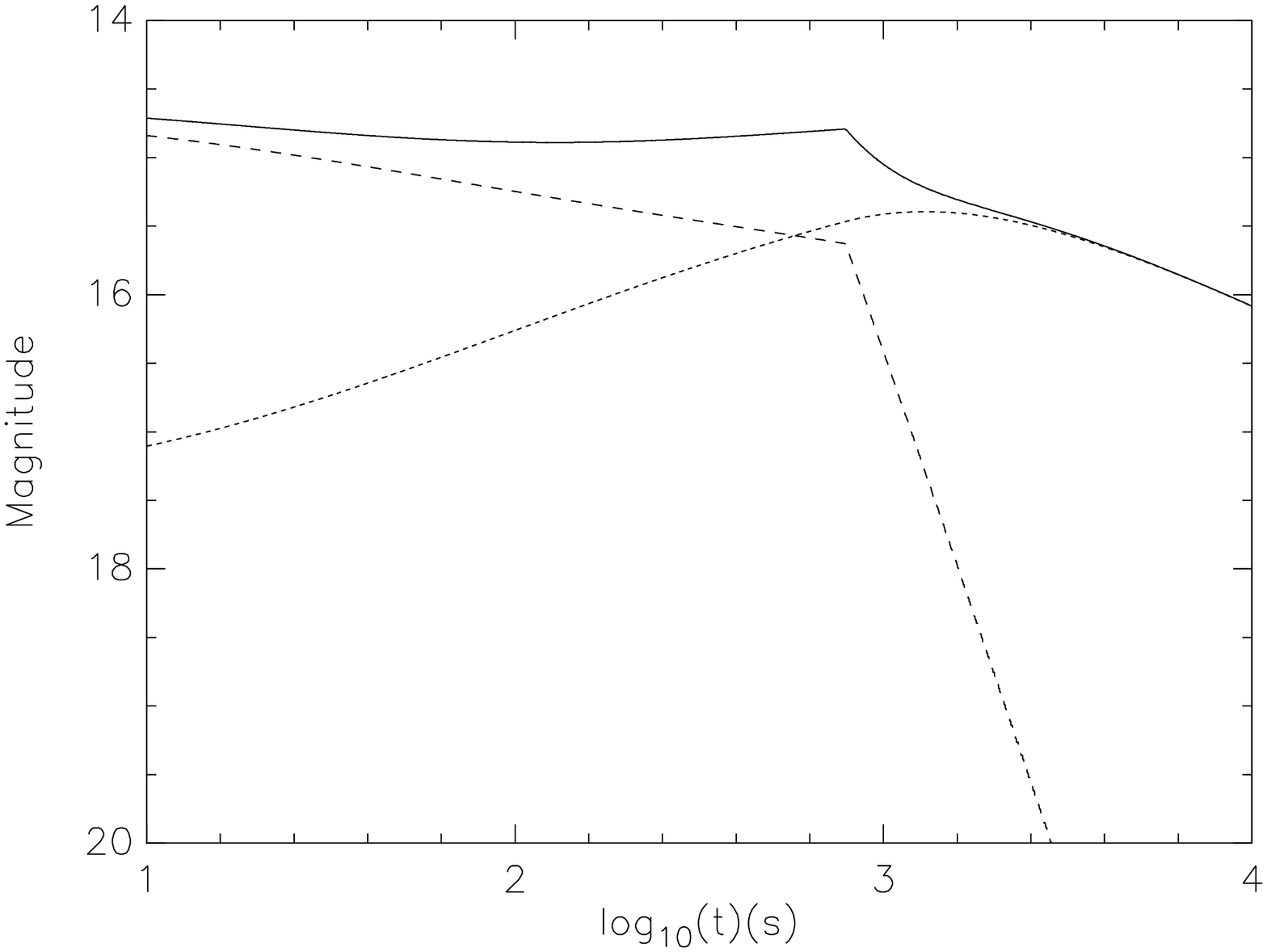}
  \caption{Optical magnitude as function of time. Parameters are the same as in Figure~\ref{fig:f_radio}.}
  \label{fig:f_optic}
\end{figure}

We take the parameters $\eta=300, E_0=1.0\times10^{52}~\mbox{erg}, A_{*}=0.1, \Delta_0=5.0\times10^{12}~\mbox{cm}, \epsilon_e=0.3, \epsilon_B=0.1$, and $D=1.0\times 10^{28}$ cm, for the RRS case. For the NRS case, we set $\eta=30$, while keeping the same other parameters as in the RRS case. Following the above analysis, we can get the emission from the two shocked regions, of which the optical magnitude at frequency $\nu = 4.55\times 10^{14}$Hz is shown in Figure \ref{fig:f_optic} [later] for RRS case and NRS case respectively. The reverse shock dominates the emission at the beginning and fades after the shock crosses the shell, which is identical for both RRS and NRS. This effect may be the cause of the so-called optical flash.

After the reverse shock crosses the shell, we choose the parameter $g=1$ for the dynamic evolution of NRS. Kobayashi and Sari (2000) discussed that $g$ should satisfy $3/2<g<7/2$ in the ISM environment. A similar conclusion can be drawn for the dynamics of the ejecta in the wind environment. As the NRS cannot decrease the velocity of the ejected shell effectively, the shocked ejecta should be quicker than the one in RRS case, which satisfies $\gamma_3\propto r^{-3/2}$. On the other hand, the ejecta must lag behind the forward shock, which satisfies $\gamma_2\propto r^{-1/2}$. So the range of $g$ should then obey $1/2<g<3/2$. What's more, the evolutions of the hydrodynamics and the emission do not depend on the value of $g$ sensitively. The evolution of $\gamma$ with the observer's time has a narrow range from $t_{\oplus}^{-1/4}$ to $t_{\oplus}^{-3/7}$ corresponding to the range of $g$.

\begin{figure}
  \centering
  \includegraphics[angle=270,width=0.4\textwidth]{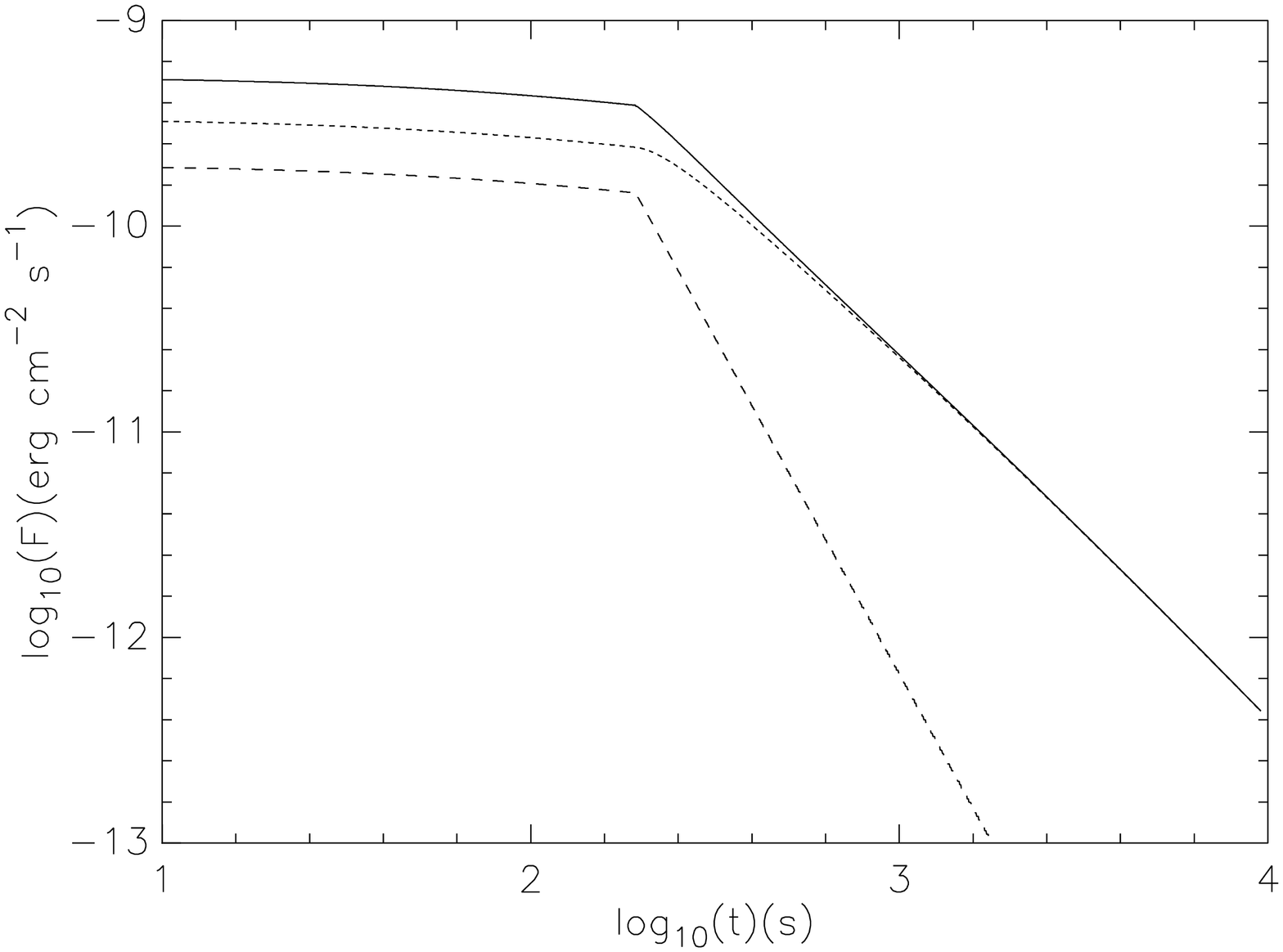}\\
  \includegraphics[angle=270,width=0.4\textwidth]{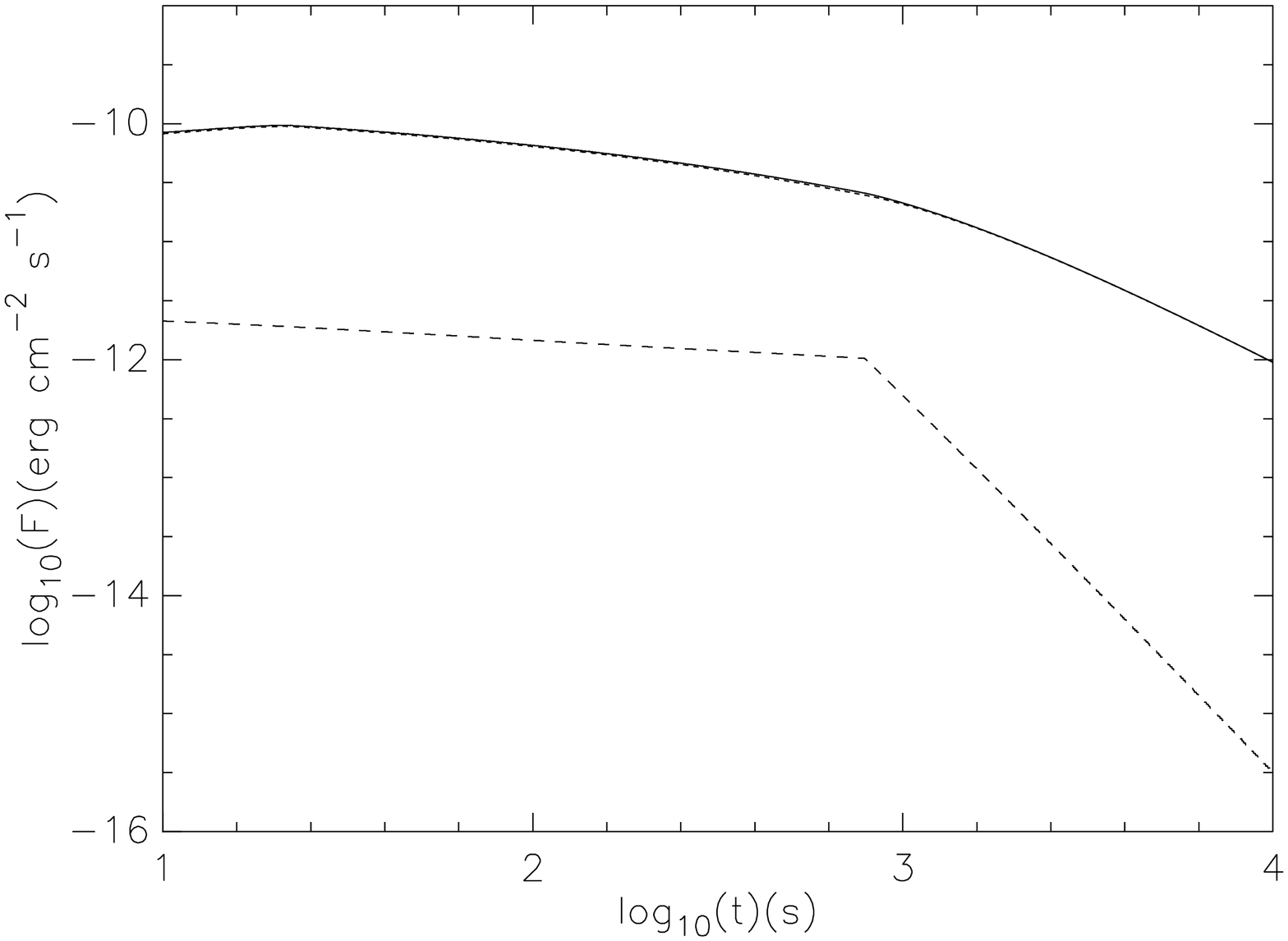}
  \caption{2-10 kev flux as function of time. Parameters are the same as in Figure~\ref{fig:f_radio}.}
  \label{fig:f_x-ray}
\end{figure}

Figures \ref{fig:f_radio}-\ref{fig:f_x-ray} show the light curves at radio (8.46 GHz), optical band ($4.55\times10^{14}$ Hz) and X-ray ($1.0\times 10^{18}$ Hz) respectively. The upper panel denotes the RRS case, and the lower panel denotes the NRS case. At low frequencies, $\nu_a$ is always greater than the observed frequency, so the emission at these frequencies is affected by the synchrotron self-absorption enormously, and can be estimated as thermal emission at this band (Chevalier \& Li 2000). The radio flux density increases with time before and shortly after the crossing time, as shown in Figure \ref{fig:f_radio}, which comes from the increasing number of the accelerated electrons. The flux will be intense enough to be detected if the distance is not so large, as the flux is inversely proportional to the square of the luminosity distance.

The numerical results are well consistent with the analytical ones. For the typical parameters and $\nu = 4.55\times 10^{14}$ Hz as the observed frequencies, at the crossing time, the orders of the typical frequencies are $\nu_{c,3} < \nu_{a,3} < \nu < \nu_{m,3}$ for RRS case, $\nu_{c,3} < \nu_{m,3} < \nu_{a,3} < \nu$ for NRS case, and $\nu_{c,2} < \nu_{a,2} < \nu < \nu_{m,2}$ for both cases. From Table \ref{tab:flux_wind}, we find that the corresponding temporal indices are $1/2$, $-1/4$ and $1/2$ for the time before the reverse shock crosses the ejected shell, where $p = 2.5$. In Figure \ref{fig:f_optic}, the slopes can be seen from the four dashed lines before the break point, which is the crossing time. The value of the flux density from region 3 at time $t = T'$ is however not consistent with the value ($55$ mJy) given by equation (\ref{eq:flux3f33_NRS}), which is about $3.5$ mJy in the figure, since the reverse shock is mildly relativistic. The curves are not accordant well with the approximated analytical slopes either. \footnote{The reverse-forward shock is assumed to begin at the fireball's coasting period, which is the initial time for the early-afterglow. However, the coasting radius is not zero, though it can be neglected at late times, which has been adopted in the scaling-law analyses. Therefore, at early times, the curve in Figure \ref{fig:f_optic} for RRS case is not straight. As the curve for NRS in the figure begins at $10$ s, the influence of the nonzero initial radius can be neglected now.}

For the optical band, the reverse shock dominates the emission at the beginning, and decays quickly after the crossing time, since there are no fresh shocked electrons to produce the emission. This is the same for both RRS and NRS cases, as seen in Figure \ref{fig:f_optic}. However, the X-ray afterglow is always dominated by the forward shock, especially for the NRS case, since the reverse shock is not strong enough, and can't accelerate the electrons to a high stochastic Lorentz factor to emit numerous X-ray band photons. Figure \ref{fig:f_x-ray} shows the emission at X-ray band for both RRS and NRS cases. From these three figures, we can see that the main emission is approximately at optical band.

\begin{figure}
  \centering
  \includegraphics[angle=270,width=0.5\textwidth]{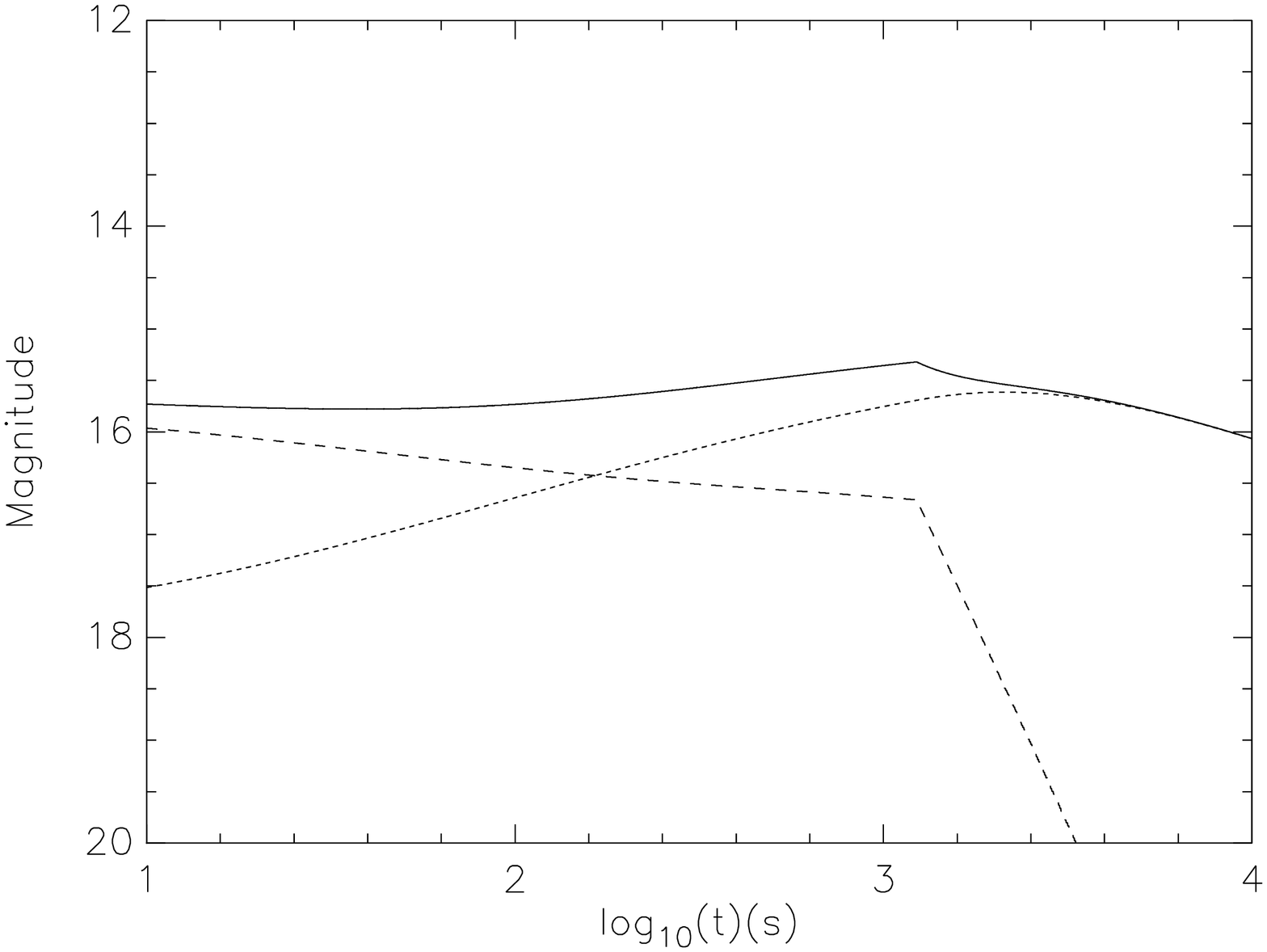}\\
  \includegraphics[angle=270,width=0.5\textwidth]{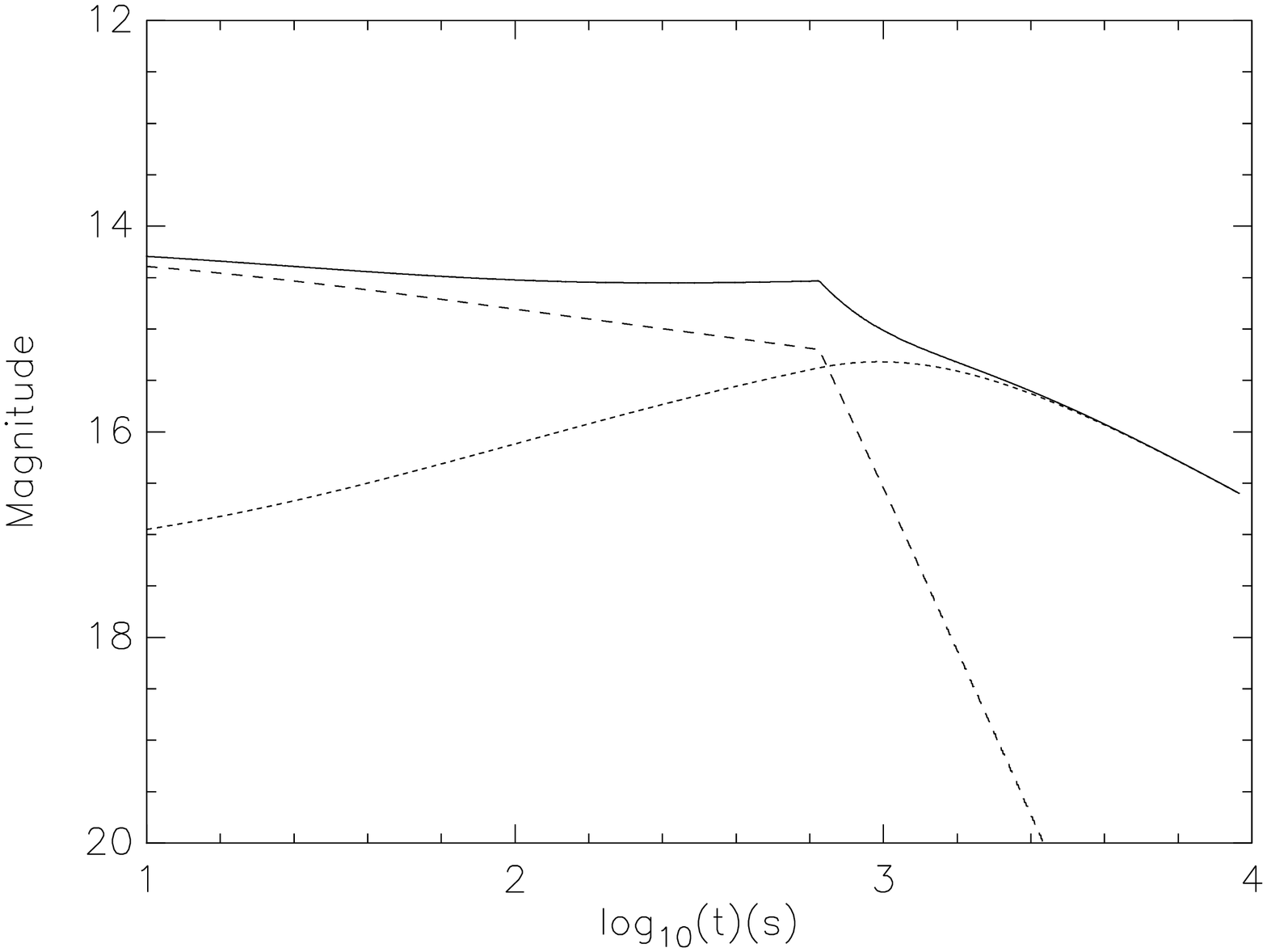}\\
  \includegraphics[angle=270,width=0.5\textwidth]{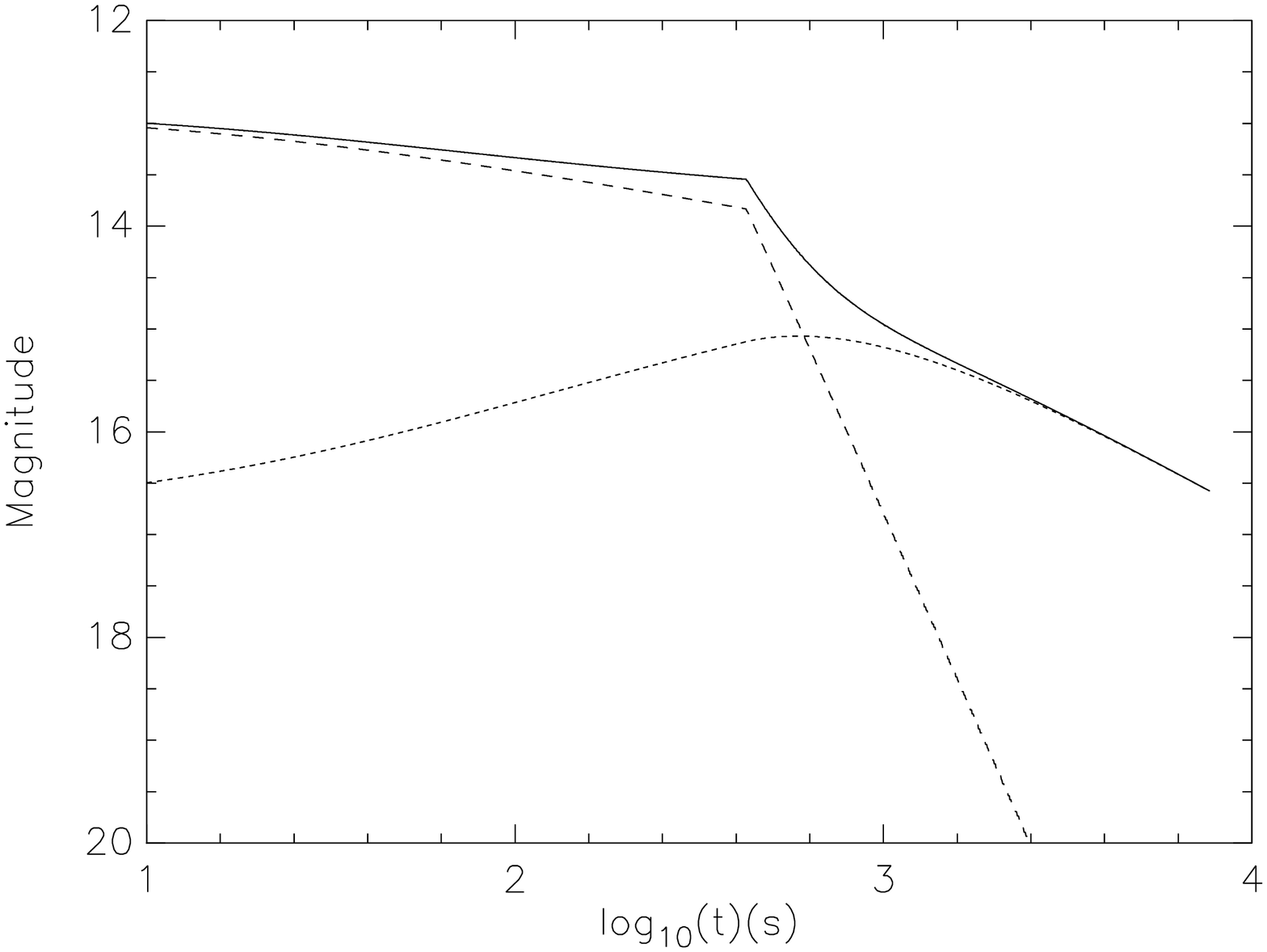}
  \caption{Magnitude as function of time at $4.55\times 10^{14}$ Hz. Parameters are the same as in Figure~\ref{fig:f_radio} except for $\eta$. The $\eta$ is $25,35$, and $50$ from top to down.}
  \label{fig:f_ETA}
\end{figure}

As seen in equation (\ref{eq:flux3f33_NRS}), the flux density depends on $\eta$ very sensitively. We plot the magnitude as function of time for different $\eta$ values in Figure \ref{fig:f_ETA}. Taking into account the lower panel in Figure \ref{fig:f_optic} and the first two in this figure, we can find that, with other parameters unchanged, the larger $\eta$, the larger the flux density, as the flux density is proportional to about $\eta^{8.5}$ if $p=2.5$. When $\eta=50$, shown in the lowest panel in Figure \ref{fig:f_ETA}, the reverse shock becomes mildly relativistic. In the relativistic reverse shock case, it is approximately inversely proportional to $\eta$. The flux density descends with the increase of $\eta$. Another phenomenon is that, with the increase of the $\eta$, the time for the emission from region 2 to overtake the one from region 3 postpones, and then region 3 almost dominate the emission during the whole early period. Because the number of the electrons in region 3 is much larger than that in region 2, the emission is dominated by region 3, when the reverse shock is powerful enough to accelerate the electrons to emit enough optical band synchrotron photons. Thirdly, no distinct ascending of the optical light curves appears before the crossing time for the NRS case. On the contrary, the light curves will descend at the beginning if the emission is dominated by the region 3.

\begin{figure}
  \centering
  \includegraphics[angle=270,width=0.5\textwidth]{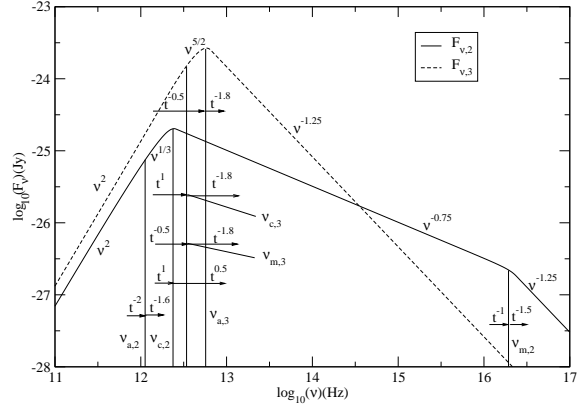}
  \caption{Flux density in region 2 (solid line) and in region 3 (dashed line) as function of the observed frequency, at the crossing time. Parameters are the same as in Figure~\ref{fig:f_radio} with $\eta=30$. The arrows on the left of the vertical line denote the time behavior of the corresponding typical frequencies before the crossing time, and the arrows on the right denote the time behavior after the crossing time.}
  \label{fig:f_nu}
\end{figure}

We plot the spectrum at the crossing time for reverse shock and forward shock respectively in Figure~\ref{fig:f_nu}. The spectrum is a typical synchrotron spectrum for the electron energy distribution with index $p=2.5$. The breaks are smoothened by the time equal arrival effect (Sari 1998). Both curves have three typical frequencies $\nu_a$, $\nu_c$ and $\nu_m$. The time behavior of the frequencies before and after the crossing time is illustrated in this figure. Before the crossing time, $\nu_{m,3}\propto t^{-0.5}$, which is different from the analytical result for NRS case $\nu_{m,3}\propto t^{1}$ (listed in Table~\ref{tab:var_wind}). This comes from the fact that the term $\bar{\gamma_3}-1$ can't be taken to be much less than 1, especially for a shallow Newtonian reverse shock. Consequently, $\nu_{a,3}$ becomes approximately $\propto t^{-0.5}$, not $\propto t^{-(p+3)/(p+5)}$ in Table~\ref{tab:var_wind}. For these parameters, $\nu_{m,3}$ is occasionally equal to $\nu_{c,3}$ at the crossing time. After the crossing time, as no fresh electrons supply, electrons with stochastic Lorentz factor greater than $\gamma_c$ disappears. The maximum electron Lorentz factor $\gamma_{cut}$ varies with time like $\gamma_m$ (Kobayashi 2000). The corresponding $\nu_{cut}$ does so. As one can see, the maximum typical frequency in region 3 is $\nu_a$, so the synchrotron self-absorption effect is important for the reverse shock in NRS case.

\begin{figure}
  \centering
  \includegraphics[angle=270,width=0.5\textwidth]{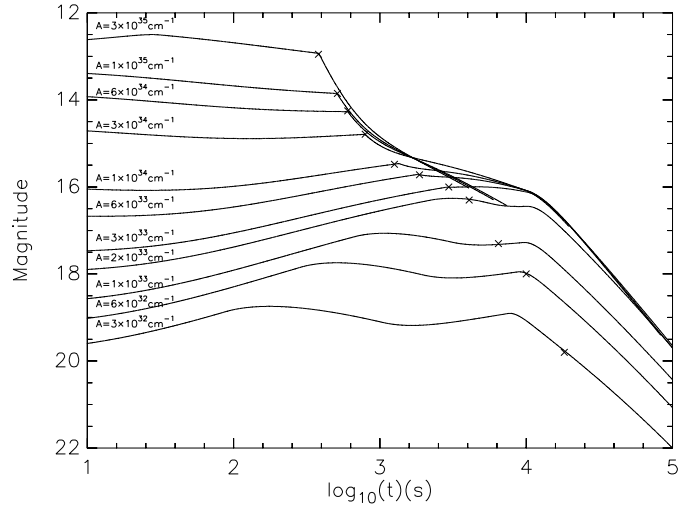}
  \caption{Light curves in optical band ($4.55\times10^{14}$Hz) for different wind parameter $A$. Parameters are the same as in Figure~\ref{fig:f_radio} except $A$, and $\eta=30$. The crosses ($\times$) indicate the crossing time of the reverse shock.}
  \label{fig:f_A}
\end{figure}

The wind parameter $A$ is important for the reverse-forward shock. How do the light curves vary if the wind density varies? We give a set of light curves of the early afterglow with different $A$ in Figure~\ref{fig:f_A}. The parameter $A$ is taken from $3\times10^{32}~\mbox{cm}^{-1}$ to $3\times10^{35}~\mbox{cm}^{-1}$ ($A_{*}=1$). With the increase of $A$, the reverse shock converts from Newtonian to relativistic. For the extreme NRS case, the emission is dominated by the forward shock (the lower light curves in Figure~\ref{fig:f_A}). This makes the light curves (summation of region 2 and region 3) have no break at the crossing time. But with the increase of $A$, the early emission are gradually dominated by the reverse shock, so the breaks (in the upper light curves) appear at the crossing time. Another phenomenon is that, as $A$ decreases, the crossing time becomes longer, which is mainly due to the spreading of the ejected shell. With the approximation by ignoring the spreading effect, the analytical crossing time for RRS is a constant $(1+z)\Delta_0/2c$. In Figure~\ref{fig:f_A}, we can see that the crossing time converges to $(1+z)\Delta_0/2c=167((1+z)/2)\Delta_{0,5\times10^{12}}$s with the increase of $A$.

\begin{figure}
  \centering
  \includegraphics[angle=270,width=0.3\textwidth]{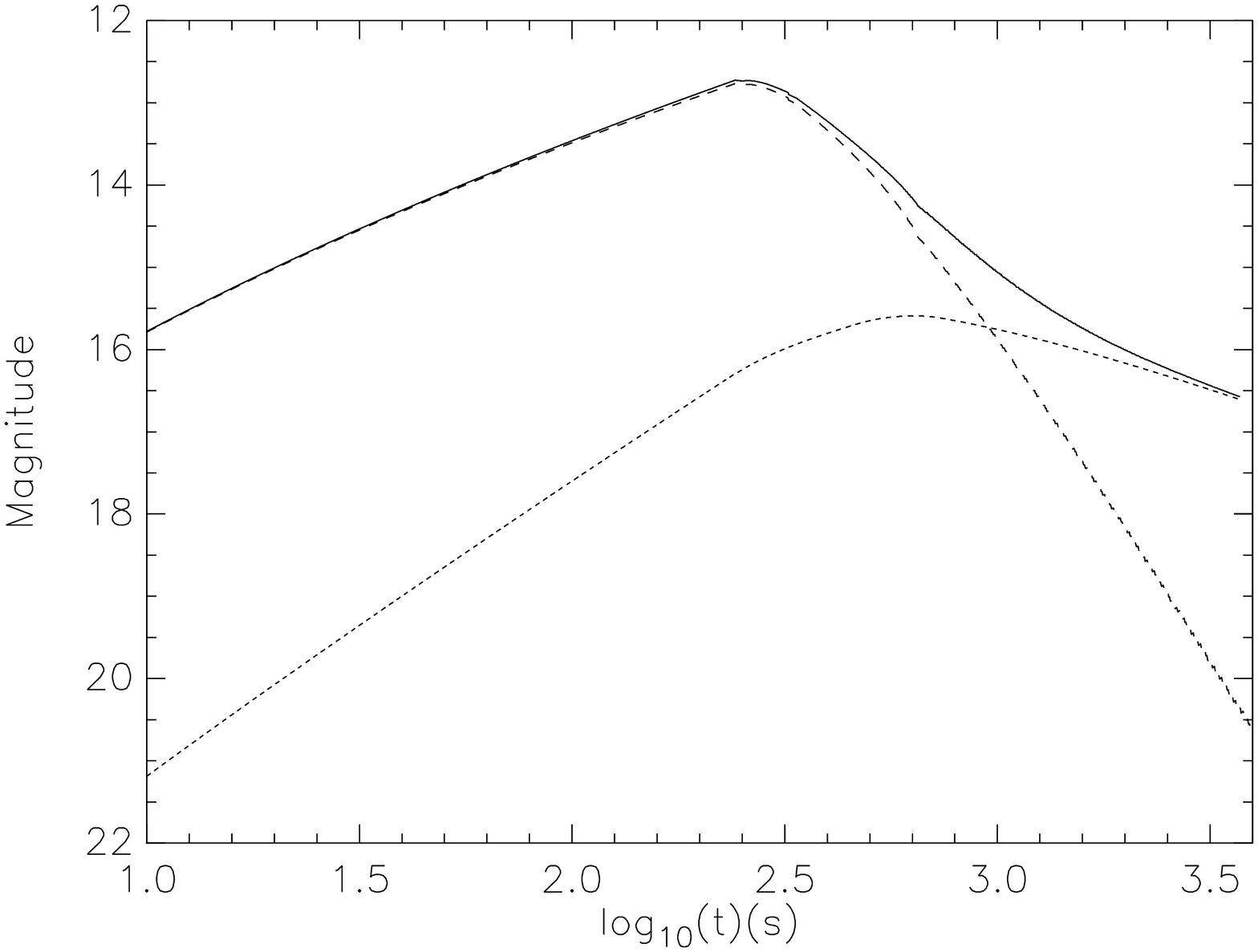}\\
  \includegraphics[angle=270,width=0.3\textwidth]{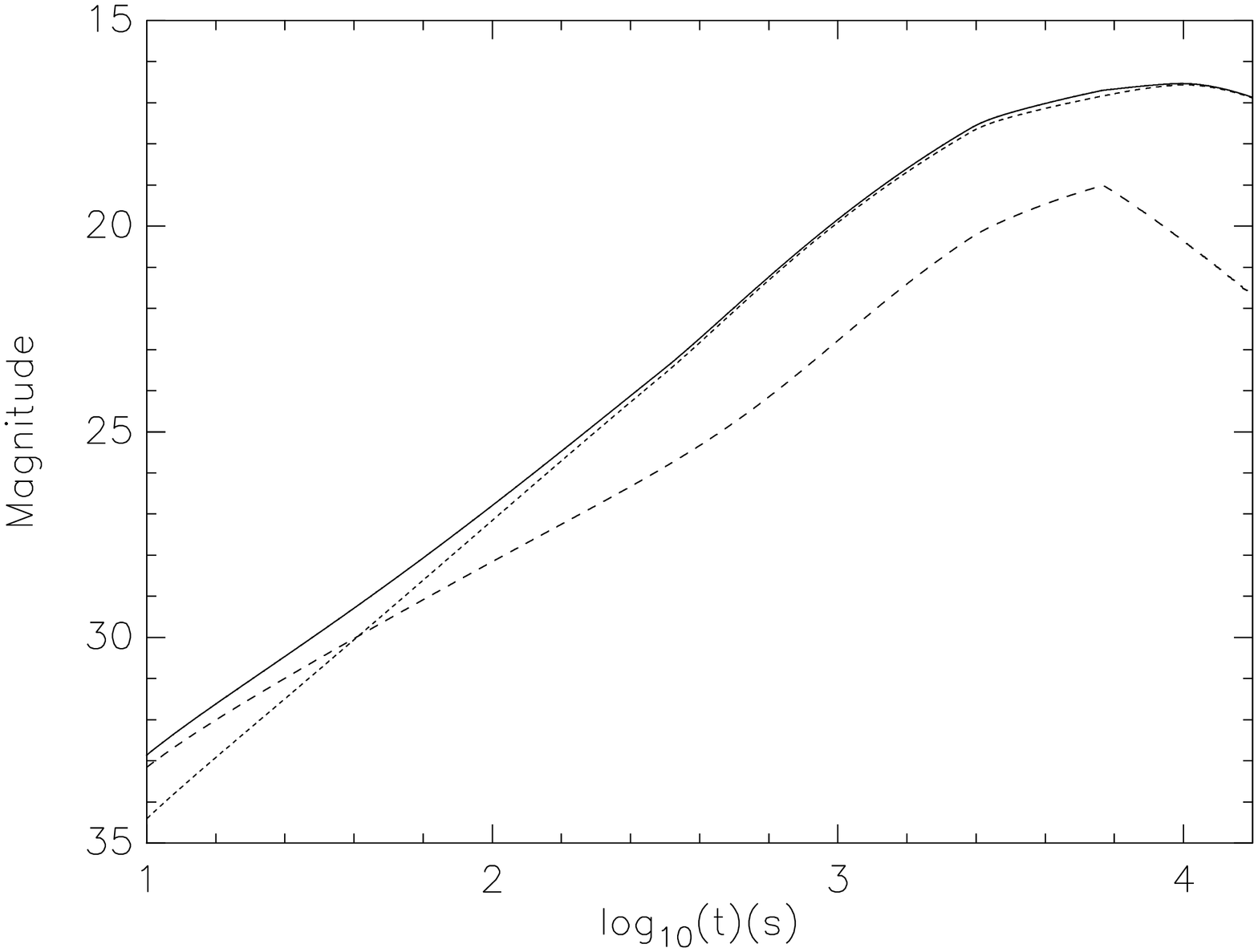}
  \caption{Magnitude at $4.55\times10^{14}$ Hz as function of time in ISM environment. $n_1=1$cm$^{-3}$ is the number density of ISM material. $\eta=300$ for the upper panel is RRS case, and $\eta=30$ for the lower panel is NRS case. Other parameters are the same as in Figure~\ref{fig:f_radio}.}
  \label{fig:f_ISM}
\end{figure}

The ISM environment case has been investigated enormously (Sari \& Piran 1995; M\'esz\'aros \& Rees 1997; Kobayashi \& Sari 2000; Kobayashi 2000; Zhang et~al. 2003). We here calculate the light curves of the reverse-forward shock for the ISM density $n_1=1$cm$^{-3}$ and with the same other parameters as the typical values in the wind environment (see Figure~\ref{fig:f_ISM}). For these parameters, the flux densities both for the RRS case (upper panel) and for the NRS case (lower panel) increase during the early period, which are different from the wind environment case. We can see in Figure~\ref{fig:f_optic} that the light curves almost keep unchanged for early times at the optical band. This may be caused by the decrease of the number density of the wind. Therefore, whether there exist a rapid increase of the early optical afterglow may be used to distinguish between the ISM and wind environments.

\section{Conclusions and Discussion} \label{sec:conclusions}

We have investigated the whole evolution of the dynamics and emission of the reverse-forward shock in a wind environment, by considering both the RRS and NRS cases. The temporal indices of the physical quantities are given in Table \ref{tab:var_wind} and \ref{tab:flux_wind}, which cover all interrelations of the typical frequencies $\nu_a$, $\nu_m$, $\nu_c$ and the observed frequency $\nu$. The flux densities of the emission at radio, optical and X-ray bands as functions of time are shown in Figures~\ref{fig:f_radio}-~\ref{fig:f_x-ray} by numerical calculations.

For the ISM model, there exists a transition radius $R_N$ (satisfies $f(R_N)=\gamma_4^2$) for thick shell (Sari \& Piran 1995), where the reverse shock becomes relativistic. An enormous difference between the wind model and the ISM model is that the Lorentz factor of the shocked regions before the crossing time is constant with time for the wind model in RRS case. Since the transition should satisfy $f/\gamma_4^2 = 1$, the ratio $f$ and $\gamma_4$ are both constant, for the wind environment, and thus no transition exists. Therefore, the relative Lorentz factor of the reverse shock $\bar{\gamma_3}$ is also independent of time. Taking into account these properties, we find that the temporal indices are relatively reliable, even if the estimates of the Lorentz factors $\bar{\gamma_3}, \gamma_2$, and $\gamma_3$ deviate from the actual values, which are caused by the Newtonian and relativistic approximations.

An optical flash emitted from the shocked shell appears for the typical parameters, but perhaps no darkening can be observed at early times. A rapid decay occurs after the reverse shock crosses the shell, and then the emission is dominated by the shocked environment material. A radio flare  lasts for a longer time. It increases continuously even shortly after the crossing time. The X-ray band emission is always dominated by region 2. At the optical band and X-ray band, the flux is nearly unchanged at early times especially for the NRS case, which may be used to diagnose the NRS in a wind environment. For the reverse shock, the synchrotron self-absorption can not be neglected, since it may exceed the other two frequencies $\nu_c$ and $\nu_m$ as the number density increases.

There is also a possibility that no prompt optical emission is detected. It may be caused by a low initial energy, a low environmental density, or strong absorption, and so on. These will decrease the flux density of the early afterglow to go beyond the detector's limits. In the {\it{Swift}}'s era, many early optical and X-ray afterglows will be detected by UVOT and XRT, like GRB050525A (Shao \& Dai 2005, Klotz et~al. 2005), and then the parameters may be determined by early afterglow data more precisely.

We would like to thank the anonymous referee for valuable suggestions. YCZ thanks T. Yan, H. L. Dai , Y. Z. Fan and Y. F. Huang for helpful discussions. This work was supported by the National Natural Science Foundation of China (grants 10233010 and 10221001), and the Ministry of Science and Technology of China (NKBRSF G19990754).

\label{lastpage}

\end{document}